\documentclass[prl,twocolumn,showpacs,superscriptaddress,floatfix]{revtex4}

\usepackage{slashed}
\usepackage{epsfig}
\usepackage{amsmath}
\usepackage{comment}
\usepackage{longtable}
\usepackage{subfigure}
\usepackage{color}

\def\met{{\mbox{$E\kern-0.57em\raise0.19ex\hbox{/}_{T}$}}}
\def\metbb{$WH,ZH\rightarrow\met b\bar{b}$}
\def\WH{$WH\rightarrow \ell\nu b\bar{b}$}

\def\lmet{$WH\rightarrow \ell\kern-0.45em\raise0.19ex\hbox{/} \nu b\bar{b}$}
\def\ZH{$ZH\rightarrow \ell^+\ell^- b\bar{b}$}

\newcommand{\METVEC}{\mbox{$\raisebox{.3ex}{$\not\!$}{\vec E}_T$}}

\newcommand{\MCFM}       {{\sc mcfm}}
\newcommand{\PYTHIA}     {{\sc pythia}}

\newcommand{\HDECAY}       {{\sc hdecay}}

\newcommand{\rar}       {\rightarrow}
\newcommand{\MET}{$\not\!\!E_T$}

\begin{document}

\title{Evidence for a particle produced in association with weak bosons and decaying to
a bottom-antibottom quark pair in Higgs boson searches at the Tevatron}

\affiliation{LAFEX, Centro Brasileiro de Pesquisas F\'{i}sicas, Rio de Janeiro, Brazil}
\affiliation{Universidade do Estado do Rio de Janeiro, Rio de Janeiro, Brazil}
\affiliation{Universidade Federal do ABC, Santo Andr\'e, Brazil}
\affiliation{Institute of Particle Physics: McGill University, Montr\'{e}al, Qu\'{e}bec, Canada H3A~2T8; Simon Fraser University, Burnaby, British Columbia, Canada V5A~1S6; University of Toronto, Toronto, Ontario, Canada M5S~1A7; and TRIUMF, Vancouver, British Columbia, Canada V6T~2A3}
\affiliation{University of Science and Technology of China, Hefei, People's Republic of China}
\affiliation{Institute of Physics, Academia Sinica, Taipei, Taiwan 11529, Republic of China}
\affiliation{Universidad de los Andes, Bogot\'a, Colombia}
\affiliation{Charles University, Faculty of Mathematics and Physics, Center for Particle Physics, Prague, Czech Republic}
\affiliation{Czech Technical University in Prague, Prague, Czech Republic}
\affiliation{Center for Particle Physics, Institute of Physics, Academy of Sciences of the Czech Republic, Prague, Czech Republic}
\affiliation{Universidad San Francisco de Quito, Quito, Ecuador}
\affiliation{Division of High Energy Physics, Department of Physics, University of Helsinki and Helsinki Institute of Physics, FIN-00014, Helsinki, Finland}
\affiliation{LPC, Universit\'e Blaise Pascal, CNRS/IN2P3, Clermont, France}
\affiliation{LPSC, Universit\'e Joseph Fourier Grenoble 1, CNRS/IN2P3, Institut National Polytechnique de Grenoble, Grenoble, France}
\affiliation{CPPM, Aix-Marseille Universit\'e, CNRS/IN2P3, Marseille, France}
\affiliation{LAL, Universit\'e Paris-Sud, CNRS/IN2P3, Orsay, France}
\affiliation{LPNHE, Universit\'es Paris VI and VII, CNRS/IN2P3, Paris, France}
\affiliation{CEA, Irfu, SPP, Saclay, France}
\affiliation{IPHC, Universit\'e de Strasbourg, CNRS/IN2P3, Strasbourg, France}
\affiliation{IPNL, Universit\'e Lyon 1, CNRS/IN2P3, Villeurbanne, France and Universit\'e de Lyon, Lyon, France}
\affiliation{III. Physikalisches Institut A, RWTH Aachen University, Aachen, Germany}
\affiliation{Physikalisches Institut, Universit\"at Freiburg, Freiburg, Germany}
\affiliation{II. Physikalisches Institut, Georg-August-Universit\"at G\"ottingen, G\"ottingen, Germany}
\affiliation{Institut f\"{u}r Experimentelle Kernphysik, Karlsruhe Institute of Technology, D-76131 Karlsruhe, Germany}
\affiliation{Institut f\"ur Physik, Universit\"at Mainz, Mainz, Germany}
\affiliation{Ludwig-Maximilians-Universit\"at M\"unchen, M\"unchen, Germany}
\affiliation{Fachbereich Physik, Bergische Universit\"at Wuppertal, Wuppertal, Germany}
\affiliation{University of Athens, 157 71 Athens, Greece}
\affiliation{Panjab University, Chandigarh, India}
\affiliation{Delhi University, Delhi, India}
\affiliation{Tata Institute of Fundamental Research, Mumbai, India}
\affiliation{University College Dublin, Dublin, Ireland}
\affiliation{Istituto Nazionale di Fisica Nucleare Bologna, $^{\sharp{a}}$University of Bologna, I-40127 Bologna, Italy}
\affiliation{Laboratori Nazionali di Frascati, Istituto Nazionale di Fisica Nucleare, I-00044 Frascati, Italy}
\affiliation{Istituto Nazionale di Fisica Nucleare, Sezione di Padova-Trento, $^{\sharp{b}}$University of Padova, I-35131 Padova, Italy}
\affiliation{Istituto Nazionale di Fisica Nucleare Pisa, $^{\sharp{c}}$University of Pisa, $^{\sharp{d}}$University of Siena and $^{\sharp{e}}$Scuola Normale Superiore, I-56127 Pisa, Italy}
\affiliation{Istituto Nazionale di Fisica Nucleare, Sezione di Roma 1, $^{\sharp{f}}$Sapienza Universit\`{a} di Roma, I-00185 Roma, Italy}
\affiliation{Istituto Nazionale di Fisica Nucleare Trieste/Udine, I-34100 Trieste, $^{\sharp{g}}$University of Udine, I-33100 Udine, Italy}
\affiliation{Okayama University, Okayama 700-8530, Japan}
\affiliation{Osaka City University, Osaka 588, Japan}
\affiliation{Waseda University, Tokyo 169, Japan}
\affiliation{University of Tsukuba, Tsukuba, Ibaraki 305, Japan}
\affiliation{Center for High Energy Physics: Kyungpook National University, Daegu 702-701, Korea; Seoul National University, Seoul 151-742, Korea; Sungkyunkwan University, Suwon 440-746, Korea; Korea Institute of Science and Technology Information, Daejeon 305-806, Korea; Chonnam National University, Gwangju 500-757, Korea; Chonbuk National University, Jeonju 561-756, Korea}
\affiliation{Korea Detector Laboratory, Korea University, Seoul, Korea}
\affiliation{CINVESTAV, Mexico City, Mexico}
\affiliation{Nikhef, Science Park, Amsterdam, the Netherlands}
\affiliation{Radboud University Nijmegen, Nijmegen, the Netherlands}
\affiliation{Joint Institute for Nuclear Research, Dubna, Russia}
\affiliation{Institution for Theoretical and Experimental Physics, ITEP, Moscow 117259, Russia}
\affiliation{Moscow State University, Moscow, Russia}
\affiliation{Institute for High Energy Physics, Protvino, Russia}
\affiliation{Petersburg Nuclear Physics Institute, St. Petersburg, Russia}
\affiliation{Comenius University, 842 48 Bratislava, Slovakia; Institute of Experimental Physics, 040 01 Kosice, Slovakia}
\affiliation{Institut de Fisica d'Altes Energies, ICREA, Universitat Autonoma de Barcelona, E-08193, Bellaterra (Barcelona), Spain}
\affiliation{Instituci\'{o} Catalana de Recerca i Estudis Avan\c{c}ats (ICREA) and Institut de F\'{i}sica d'Altes Energies (IFAE), Barcelona, Spain}
\affiliation{Centro de Investigaciones Energeticas Medioambientales y Tecnologicas, E-28040 Madrid, Spain}
\affiliation{Instituto de Fisica de Cantabria, CSIC-University of Cantabria, 39005 Santander, Spain}
\affiliation{Uppsala University, Uppsala, Sweden}
\affiliation{University of Geneva, CH-1211 Geneva 4, Switzerland}
\affiliation{Glasgow University, Glasgow G12 8QQ, United Kingdom}
\affiliation{Lancaster University, Lancaster LA1 4YB, United Kingdom}
\affiliation{University of Liverpool, Liverpool L69 7ZE, United Kingdom}
\affiliation{Imperial College London, London SW7 2AZ, United Kingdom}
\affiliation{University College London, London WC1E 6BT, United Kingdom}
\affiliation{The University of Manchester, Manchester M13 9PL, United Kingdom}
\affiliation{University of Oxford, Oxford OX1 3RH, United Kingdom}
\affiliation{University of Arizona, Tucson, Arizona 85721, USA}
\affiliation{Ernest Orlando Lawrence Berkeley National Laboratory, Berkeley, California 94720, USA}
\affiliation{University of California, Davis, Davis, California 95616, USA}
\affiliation{University of California, Los Angeles, Los Angeles, California 90024, USA}
\affiliation{University of California Riverside, Riverside, California 92521, USA}
\affiliation{Yale University, New Haven, Connecticut 06520, USA}
\affiliation{University of Florida, Gainesville, Florida 32611, USA}
\affiliation{Florida State University, Tallahassee, Florida 32306, USA}
\affiliation{Argonne National Laboratory, Argonne, Illinois 60439, USA}
\affiliation{Fermi National Accelerator Laboratory, Batavia, Illinois 60510, USA}
\affiliation{Enrico Fermi Institute, University of Chicago, Chicago, Illinois 60637, USA}
\affiliation{University of Illinois at Chicago, Chicago, Illinois 60607, USA}
\affiliation{Northern Illinois University, DeKalb, Illinois 60115, USA}
\affiliation{Northwestern University, Evanston, Illinois 60208, USA}
\affiliation{University of Illinois, Urbana, Illinois 61801, USA}
\affiliation{Indiana University, Bloomington, Indiana 47405, USA}
\affiliation{Purdue University Calumet, Hammond, Indiana 46323, USA}
\affiliation{University of Notre Dame, Notre Dame, Indiana 46556, USA}
\affiliation{Purdue University, West Lafayette, Indiana 47907, USA}
\affiliation{Iowa State University, Ames, Iowa 50011, USA}
\affiliation{University of Kansas, Lawrence, Kansas 66045, USA}
\affiliation{Kansas State University, Manhattan, Kansas 66506, USA}
\affiliation{Louisiana Tech University, Ruston, Louisiana 71272, USA}
\affiliation{The Johns Hopkins University, Baltimore, Maryland 21218, USA}
\affiliation{Boston University, Boston, Massachusetts 02215, USA}
\affiliation{Northeastern University, Boston, Massachusetts 02115, USA}
\affiliation{Harvard University, Cambridge, Massachusetts 02138, USA}
\affiliation{Massachusetts Institute of Technology, Cambridge, Massachusetts 02139, USA}
\affiliation{Tufts University, Medford, Massachusetts 02155, USA}
\affiliation{University of Michigan, Ann Arbor, Michigan 48109, USA}
\affiliation{Wayne State University, Detroit, Michigan 48201, USA}
\affiliation{Michigan State University, East Lansing, Michigan 48824, USA}
\affiliation{University of Mississippi, University, Mississippi 38677, USA}
\affiliation{University of Nebraska, Lincoln, Nebraska 68588, USA}
\affiliation{Rutgers University, Piscataway, New Jersey 08855, USA}
\affiliation{Princeton University, Princeton, New Jersey 08544, USA}
\affiliation{University of New Mexico, Albuquerque, New Mexico 87131, USA}
\affiliation{State University of New York, Buffalo, New York 14260, USA}
\affiliation{The Rockefeller University, New York, New York 10065, USA}
\affiliation{University of Rochester, Rochester, New York 14627, USA}
\affiliation{State University of New York, Stony Brook, New York 11794, USA}
\affiliation{Brookhaven National Laboratory, Upton, New York 11973, USA}
\affiliation{Duke University, Durham, North Carolina 27708, USA}
\affiliation{The Ohio State University, Columbus, Ohio 43210, USA}
\affiliation{Langston University, Langston, Oklahoma 73050, USA}
\affiliation{University of Oklahoma, Norman, Oklahoma 73019, USA}
\affiliation{Oklahoma State University, Stillwater, Oklahoma 74078, USA}
\affiliation{University of Pennsylvania, Philadelphia, Pennsylvania 19104, USA}
\affiliation{Carnegie Mellon University, Pittsburgh, Pennsylvania 15213, USA}
\affiliation{University of Pittsburgh, Pittsburgh, Pennsylvania 15260, USA}
\affiliation{Brown University, Providence, Rhode Island 02912, USA}
\affiliation{University of Texas, Arlington, Texas 76019, USA}
\affiliation{Texas A\&M University, College Station, Texas 77843, USA}
\affiliation{Southern Methodist University, Dallas, Texas 75275, USA}
\affiliation{Rice University, Houston, Texas 77005, USA}
\affiliation{Baylor University, Waco, Texas 76798, USA}
\affiliation{University of Virginia, Charlottesville, Virginia 22904, USA}
\affiliation{University of Washington, Seattle, Washington 98195, USA}
\affiliation{University of Wisconsin, Madison, Wisconsin 53706, USA}
\author{T.~Aaltonen$^{\dag}$}~\affiliation{Division of High Energy Physics, Department of Physics, University of Helsinki and Helsinki Institute of Physics, FIN-00014, Helsinki, Finland}
\author{V.M.~Abazov$^{\ddag}$}~\affiliation{Joint Institute for Nuclear Research, Dubna, Russia}
\author{B.~Abbott$^{\ddag}$}~\affiliation{University of Oklahoma, Norman, Oklahoma 73019, USA}
\author{B.S.~Acharya$^{\ddag}$}~\affiliation{Tata Institute of Fundamental Research, Mumbai, India}
\author{M.~Adams$^{\ddag}$}~\affiliation{University of Illinois at Chicago, Chicago, Illinois 60607, USA}
\author{T.~Adams$^{\ddag}$}~\affiliation{Florida State University, Tallahassee, Florida 32306, USA}
\author{G.D.~Alexeev$^{\ddag}$}~\affiliation{Joint Institute for Nuclear Research, Dubna, Russia}
\author{G.~Alkhazov$^{\ddag}$}~\affiliation{Petersburg Nuclear Physics Institute, St. Petersburg, Russia}
\author{A.~Alton$^{\ddag a}$}~\affiliation{University of Michigan, Ann Arbor, Michigan 48109, USA}
\author{B.~\'{A}lvarez~Gonz\'{a}lez$^{\dag a}$}~\affiliation{Instituto de Fisica de Cantabria, CSIC-University of Cantabria, 39005 Santander, Spain}
\author{G.~Alverson$^{\ddag}$}~\affiliation{Northeastern University, Boston, Massachusetts 02115, USA}
\author{S.~Amerio$^{\dag}$}~\affiliation{Istituto Nazionale di Fisica Nucleare, Sezione di Padova-Trento, $^{\sharp{b}}$University of Padova, I-35131 Padova, Italy}
\author{D.~Amidei$^{\dag}$}~\affiliation{University of Michigan, Ann Arbor, Michigan 48109, USA}
\author{A.~Anastassov$^{\dag b}$}~\affiliation{Fermi National Accelerator Laboratory, Batavia, Illinois 60510, USA}
\author{A.~Annovi$^{\dag}$}~\affiliation{Laboratori Nazionali di Frascati, Istituto Nazionale di Fisica Nucleare, I-00044 Frascati, Italy}
\author{J.~Antos$^{\dag}$}~\affiliation{Comenius University, 842 48 Bratislava, Slovakia; Institute of Experimental Physics, 040 01 Kosice, Slovakia}
\author{G.~Apollinari$^{\dag}$}~\affiliation{Fermi National Accelerator Laboratory, Batavia, Illinois 60510, USA}
\author{J.A.~Appel$^{\dag}$}~\affiliation{Fermi National Accelerator Laboratory, Batavia, Illinois 60510, USA}
\author{T.~Arisawa$^{\dag}$}~\affiliation{Waseda University, Tokyo 169, Japan}
\author{A.~Artikov$^{\dag}$}~\affiliation{Joint Institute for Nuclear Research, Dubna, Russia}
\author{J.~Asaadi$^{\dag}$}~\affiliation{Texas A\&M University, College Station, Texas 77843, USA}
\author{W.~Ashmanskas$^{\dag}$}~\affiliation{Fermi National Accelerator Laboratory, Batavia, Illinois 60510, USA}
\author{A.~Askew$^{\ddag}$}~\affiliation{Florida State University, Tallahassee, Florida 32306, USA}
\author{S.~Atkins$^{\ddag}$}~\affiliation{Louisiana Tech University, Ruston, Louisiana 71272, USA}
\author{B.~Auerbach$^{\dag}$}~\affiliation{Yale University, New Haven, Connecticut 06520, USA}
\author{K.~Augsten$^{\ddag}$}~\affiliation{Czech Technical University in Prague, Prague, Czech Republic}
\author{A.~Aurisano$^{\dag}$}~\affiliation{Texas A\&M University, College Station, Texas 77843, USA}
\author{C.~Avila$^{\ddag}$}~\affiliation{Universidad de los Andes, Bogot\'a, Colombia}
\author{F.~Azfar$^{\dag}$}~\affiliation{University of Oxford, Oxford OX1 3RH, United Kingdom}
\author{F.~Badaud$^{\ddag}$}~\affiliation{LPC, Universit\'e Blaise Pascal, CNRS/IN2P3, Clermont, France}
\author{W.~Badgett$^{\dag}$}~\affiliation{Fermi National Accelerator Laboratory, Batavia, Illinois 60510, USA}
\author{T.~Bae$^{\dag}$}~\affiliation{Center for High Energy Physics: Kyungpook National University, Daegu 702-701, Korea; Seoul National University, Seoul 151-742, Korea; Sungkyunkwan University, Suwon 440-746, Korea; Korea Institute of Science and Technology Information, Daejeon 305-806, Korea; Chonnam National University, Gwangju 500-757, Korea; Chonbuk National University, Jeonju 561-756, Korea}
\author{L.~Bagby$^{\ddag}$}~\affiliation{Fermi National Accelerator Laboratory, Batavia, Illinois 60510, USA}
\author{B.~Baldin$^{\ddag}$}~\affiliation{Fermi National Accelerator Laboratory, Batavia, Illinois 60510, USA}
\author{D.V.~Bandurin$^{\ddag}$}~\affiliation{Florida State University, Tallahassee, Florida 32306, USA}
\author{S.~Banerjee$^{\ddag}$}~\affiliation{Tata Institute of Fundamental Research, Mumbai, India}
\author{A.~Barbaro-Galtieri$^{\dag}$}~\affiliation{Ernest Orlando Lawrence Berkeley National Laboratory, Berkeley, California 94720, USA}
\author{E.~Barberis$^{\ddag}$}~\affiliation{Northeastern University, Boston, Massachusetts 02115, USA}
\author{P.~Baringer$^{\ddag}$}~\affiliation{University of Kansas, Lawrence, Kansas 66045, USA}
\author{V.E.~Barnes$^{\dag}$}~\affiliation{Purdue University, West Lafayette, Indiana 47907, USA}
\author{B.A.~Barnett$^{\dag}$}~\affiliation{The Johns Hopkins University, Baltimore, Maryland 21218, USA}
\author{P.~Barria$^{\dag\sharp{d}}$}~\affiliation{Istituto Nazionale di Fisica Nucleare Pisa, $^{\sharp{c}}$University of Pisa, $^{\sharp{d}}$University of Siena and $^{\sharp{e}}$Scuola Normale Superiore, I-56127 Pisa, Italy}
\author{J.F.~Bartlett$^{\ddag}$}~\affiliation{Fermi National Accelerator Laboratory, Batavia, Illinois 60510, USA}
\author{P.~Bartos$^{\dag}$}~\affiliation{Comenius University, 842 48 Bratislava, Slovakia; Institute of Experimental Physics, 040 01 Kosice, Slovakia}
\author{U.~Bassler$^{\ddag}$}~\affiliation{CEA, Irfu, SPP, Saclay, France}
\author{M.~Bauce$^{\dag\sharp{b}}$}~\affiliation{Istituto Nazionale di Fisica Nucleare, Sezione di Padova-Trento, $^{\sharp{b}}$University of Padova, I-35131 Padova, Italy}
\author{V.~Bazterra$^{\ddag}$}~\affiliation{University of Illinois at Chicago, Chicago, Illinois 60607, USA}
\author{A.~Bean$^{\ddag}$}~\affiliation{University of Kansas, Lawrence, Kansas 66045, USA}
\author{F.~Bedeschi$^{\dag}$}~\affiliation{Istituto Nazionale di Fisica Nucleare Pisa, $^{\sharp{c}}$University of Pisa, $^{\sharp{d}}$University of Siena and $^{\sharp{e}}$Scuola Normale Superiore, I-56127 Pisa, Italy}
\author{M.~Begalli$^{\ddag}$}~\affiliation{Universidade do Estado do Rio de Janeiro, Rio de Janeiro, Brazil}
\author{S.~Behari$^{\dag}$}~\affiliation{The Johns Hopkins University, Baltimore, Maryland 21218, USA}
\author{L.~Bellantoni$^{\ddag}$}~\affiliation{Fermi National Accelerator Laboratory, Batavia, Illinois 60510, USA}
\author{G.~Bellettini$^{\dag\sharp{c}}$}~\affiliation{Istituto Nazionale di Fisica Nucleare Pisa, $^{\sharp{c}}$University of Pisa, $^{\sharp{d}}$University of Siena and $^{\sharp{e}}$Scuola Normale Superiore, I-56127 Pisa, Italy}
\author{J.~Bellinger$^{\dag}$}~\affiliation{University of Wisconsin, Madison, Wisconsin 53706, USA}
\author{D.~Benjamin$^{\dag}$}~\affiliation{Duke University, Durham, North Carolina 27708, USA}
\author{A.~Beretvas$^{\dag}$}~\affiliation{Fermi National Accelerator Laboratory, Batavia, Illinois 60510, USA}
\author{S.B.~Beri$^{\ddag}$}~\affiliation{Panjab University, Chandigarh, India}
\author{G.~Bernardi$^{\ddag}$}~\affiliation{LPNHE, Universit\'es Paris VI and VII, CNRS/IN2P3, Paris, France}
\author{R.~Bernhard$^{\ddag}$}~\affiliation{Physikalisches Institut, Universit\"at Freiburg, Freiburg, Germany}
\author{I.~Bertram$^{\ddag}$}~\affiliation{Lancaster University, Lancaster LA1 4YB, United Kingdom}
\author{M.~Besan\c{c}on$^{\ddag}$}~\affiliation{CEA, Irfu, SPP, Saclay, France}
\author{R.~Beuselinck$^{\ddag}$}~\affiliation{Imperial College London, London SW7 2AZ, United Kingdom}
\author{P.C.~Bhat$^{\ddag}$}~\affiliation{Fermi National Accelerator Laboratory, Batavia, Illinois 60510, USA}
\author{S.~Bhatia$^{\ddag}$}~\affiliation{University of Mississippi, University, Mississippi 38677, USA}
\author{V.~Bhatnagar$^{\ddag}$}~\affiliation{Panjab University, Chandigarh, India}
\author{A.~Bhatti$^{\dag}$}~\affiliation{The Rockefeller University, New York, New York 10065, USA}
\author{M.~Binkley$^*$$^{\dag}$}~\affiliation{Fermi National Accelerator Laboratory, Batavia, Illinois 60510, USA}
\author{D.~Bisello$^{\dag\sharp{b}}$}~\affiliation{Istituto Nazionale di Fisica Nucleare, Sezione di Padova-Trento, $^{\sharp{b}}$University of Padova, I-35131 Padova, Italy}
\author{I.~Bizjak$^{\dag}$}~\affiliation{University College London, London WC1E 6BT, United Kingdom}
\author{K.R.~Bland$^{\dag}$}~\affiliation{Baylor University, Waco, Texas 76798, USA}
\author{G.~Blazey$^{\ddag}$}~\affiliation{Northern Illinois University, DeKalb, Illinois 60115, USA}
\author{S.~Blessing$^{\ddag}$}~\affiliation{Florida State University, Tallahassee, Florida 32306, USA}
\author{K.~Bloom$^{\ddag}$}~\affiliation{University of Nebraska, Lincoln, Nebraska 68588, USA}
\author{B.~Blumenfeld$^{\dag}$}~\affiliation{The Johns Hopkins University, Baltimore, Maryland 21218, USA}
\author{A.~Bocci$^{\dag}$}~\affiliation{Duke University, Durham, North Carolina 27708, USA}
\author{A.~Bodek$^{\dag}$}~\affiliation{University of Rochester, Rochester, New York 14627, USA}
\author{A.~Boehnlein$^{\ddag}$}~\affiliation{Fermi National Accelerator Laboratory, Batavia, Illinois 60510, USA}
\author{D.~Boline$^{\ddag}$}~\affiliation{State University of New York, Stony Brook, New York 11794, USA}
\author{E.E.~Boos$^{\ddag}$}~\affiliation{Moscow State University, Moscow, Russia}
\author{G.~Borissov$^{\ddag}$}~\affiliation{Lancaster University, Lancaster LA1 4YB, United Kingdom}
\author{D.~Bortoletto$^{\dag}$}~\affiliation{Purdue University, West Lafayette, Indiana 47907, USA}
\author{T.~Bose$^{\ddag}$}~\affiliation{Boston University, Boston, Massachusetts 02215, USA}
\author{J.~Boudreau$^{\dag}$}~\affiliation{University of Pittsburgh, Pittsburgh, Pennsylvania 15260, USA}
\author{A.~Boveia$^{\dag}$}~\affiliation{Enrico Fermi Institute, University of Chicago, Chicago, Illinois 60637, USA}
\author{A.~Brandt$^{\ddag}$}~\affiliation{University of Texas, Arlington, Texas 76019, USA}
\author{O.~Brandt$^{\ddag}$}~\affiliation{II. Physikalisches Institut, Georg-August-Universit\"at G\"ottingen, G\"ottingen, Germany}
\author{L.~Brigliadori$^{\dag\sharp{a}}$}~\affiliation{Istituto Nazionale di Fisica Nucleare Bologna, $^{\sharp{a}}$University of Bologna, I-40127 Bologna, Italy}
\author{R.~Brock$^{\ddag}$}~\affiliation{Michigan State University, East Lansing, Michigan 48824, USA}
\author{C.~Bromberg$^{\dag}$}~\affiliation{Michigan State University, East Lansing, Michigan 48824, USA}
\author{A.~Bross$^{\ddag}$}~\affiliation{Fermi National Accelerator Laboratory, Batavia, Illinois 60510, USA}
\author{D.~Brown$^{\ddag}$}~\affiliation{LPNHE, Universit\'es Paris VI and VII, CNRS/IN2P3, Paris, France}
\author{J.~Brown$^{\ddag}$}~\affiliation{LPNHE, Universit\'es Paris VI and VII, CNRS/IN2P3, Paris, France}
\author{E.~Brucken$^{\dag}$}~\affiliation{Division of High Energy Physics, Department of Physics, University of Helsinki and Helsinki Institute of Physics, FIN-00014, Helsinki, Finland}
\author{J.~Budagov$^{\dag}$}~\affiliation{Joint Institute for Nuclear Research, Dubna, Russia}
\author{X.B.~Bu$^{\ddag}$}~\affiliation{Fermi National Accelerator Laboratory, Batavia, Illinois 60510, USA}
\author{H.S.~Budd$^{\dag}$}~\affiliation{University of Rochester, Rochester, New York 14627, USA}
\author{M.~Buehler$^{\ddag}$}~\affiliation{Fermi National Accelerator Laboratory, Batavia, Illinois 60510, USA}
\author{V.~Buescher$^{\ddag}$}~\affiliation{Institut f\"ur Physik, Universit\"at Mainz, Mainz, Germany}
\author{V.~Bunichev$^{\ddag}$}~\affiliation{Moscow State University, Moscow, Russia}
\author{S.~Burdin$^{\ddag b}$}~\affiliation{Lancaster University, Lancaster LA1 4YB, United Kingdom}
\author{K.~Burkett$^{\dag}$}~\affiliation{Fermi National Accelerator Laboratory, Batavia, Illinois 60510, USA}
\author{G.~Busetto$^{\dag\sharp{b}}$}~\affiliation{Istituto Nazionale di Fisica Nucleare, Sezione di Padova-Trento, $^{\sharp{b}}$University of Padova, I-35131 Padova, Italy}
\author{P.~Bussey$^{\dag}$}~\affiliation{Glasgow University, Glasgow G12 8QQ, United Kingdom}
\author{C.P.~Buszello$^{\ddag}$}~\affiliation{Uppsala University, Uppsala, Sweden}
\author{A.~Buzatu$^{\dag}$}~\affiliation{Institute of Particle Physics: McGill University, Montr\'{e}al, Qu\'{e}bec, Canada H3A~2T8; Simon Fraser University, Burnaby, British Columbia, Canada V5A~1S6; University of Toronto, Toronto, Ontario, Canada M5S~1A7; and TRIUMF, Vancouver, British Columbia, Canada V6T~2A3}
\author{A.~Calamba$^{\dag}$}~\affiliation{Carnegie Mellon University, Pittsburgh, Pennsylvania 15213, USA}
\author{C.~Calancha$^{\dag}$}~\affiliation{Centro de Investigaciones Energeticas Medioambientales y Tecnologicas, E-28040 Madrid, Spain}
\author{E.~Camacho-P\'erez$^{\ddag}$}~\affiliation{CINVESTAV, Mexico City, Mexico}
\author{S.~Camarda$^{\dag}$}~\affiliation{Institut de Fisica d'Altes Energies, ICREA, Universitat Autonoma de Barcelona, E-08193, Bellaterra (Barcelona), Spain}
\author{M.~Campanelli$^{\dag}$}~\affiliation{University College London, London WC1E 6BT, United Kingdom}
\author{M.~Campbell$^{\dag}$}~\affiliation{University of Michigan, Ann Arbor, Michigan 48109, USA}
\author{F.~Canelli$^{\dag}$}~\affiliation{Enrico Fermi Institute, University of Chicago, Chicago, Illinois 60637, USA}
\author{B.~Carls$^{\dag}$}~\affiliation{University of Illinois, Urbana, Illinois 61801, USA}
\author{D.~Carlsmith$^{\dag}$}~\affiliation{University of Wisconsin, Madison, Wisconsin 53706, USA}
\author{R.~Carosi$^{\dag}$}~\affiliation{Istituto Nazionale di Fisica Nucleare Pisa, $^{\sharp{c}}$University of Pisa, $^{\sharp{d}}$University of Siena and $^{\sharp{e}}$Scuola Normale Superiore, I-56127 Pisa, Italy}
\author{S.~Carrillo$^{\dag c}$}~\affiliation{University of Florida, Gainesville, Florida 32611, USA}
\author{S.~Carron$^{\dag}$}~\affiliation{Fermi National Accelerator Laboratory, Batavia, Illinois 60510, USA}
\author{B.~Casal$^{\dag d}$}~\affiliation{Instituto de Fisica de Cantabria, CSIC-University of Cantabria, 39005 Santander, Spain}
\author{M.~Casarsa$^{\dag}$}~\affiliation{Istituto Nazionale di Fisica Nucleare Trieste/Udine, I-34100 Trieste, $^{\sharp{g}}$University of Udine, I-33100 Udine, Italy}
\author{B.C.K.~Casey$^{\ddag}$}~\affiliation{Fermi National Accelerator Laboratory, Batavia, Illinois 60510, USA}
\author{H.~Castilla-Valdez$^{\ddag}$}~\affiliation{CINVESTAV, Mexico City, Mexico}
\author{A.~Castro$^{\dag\sharp{a}}$}~\affiliation{Istituto Nazionale di Fisica Nucleare Bologna, $^{\sharp{a}}$University of Bologna, I-40127 Bologna, Italy}
\author{P.~Catastini$^{\dag}$}~\affiliation{Harvard University, Cambridge, Massachusetts 02138, USA}
\author{S.~Caughron$^{\ddag}$}~\affiliation{Michigan State University, East Lansing, Michigan 48824, USA}
\author{D.~Cauz$^{\dag}$}~\affiliation{Istituto Nazionale di Fisica Nucleare Trieste/Udine, I-34100 Trieste, $^{\sharp{g}}$University of Udine, I-33100 Udine, Italy}
\author{V.~Cavaliere$^{\dag}$}~\affiliation{University of Illinois, Urbana, Illinois 61801, USA}
\author{M.~Cavalli-Sforza$^{\dag}$}~\affiliation{Institut de Fisica d'Altes Energies, ICREA, Universitat Autonoma de Barcelona, E-08193, Bellaterra (Barcelona), Spain}
\author{A.~Cerri$^{\dag e}$}~\affiliation{Ernest Orlando Lawrence Berkeley National Laboratory, Berkeley, California 94720, USA}
\author{L.~Cerrito$^{\dag f}$}~\affiliation{University College London, London WC1E 6BT, United Kingdom}
\author{S.~Chakrabarti$^{\ddag}$}~\affiliation{State University of New York, Stony Brook, New York 11794, USA}
\author{D.~Chakraborty$^{\ddag}$}~\affiliation{Northern Illinois University, DeKalb, Illinois 60115, USA}
\author{K.M.~Chan$^{\ddag}$}~\affiliation{University of Notre Dame, Notre Dame, Indiana 46556, USA}
\author{A.~Chandra$^{\ddag}$}~\affiliation{Rice University, Houston, Texas 77005, USA}
\author{E.~Chapon$^{\ddag}$}~\affiliation{CEA, Irfu, SPP, Saclay, France}
\author{G.~Chen$^{\ddag}$}~\affiliation{University of Kansas, Lawrence, Kansas 66045, USA}
\author{Y.C.~Chen$^{\dag}$}~\affiliation{Institute of Physics, Academia Sinica, Taipei, Taiwan 11529, Republic of China}
\author{M.~Chertok$^{\dag}$}~\affiliation{University of California, Davis, Davis, California 95616, USA}
\author{S.~Chevalier-Th\'ery$^{\ddag}$}~\affiliation{CEA, Irfu, SPP, Saclay, France}
\author{G.~Chiarelli$^{\dag}$}~\affiliation{Istituto Nazionale di Fisica Nucleare Pisa, $^{\sharp{c}}$University of Pisa, $^{\sharp{d}}$University of Siena and $^{\sharp{e}}$Scuola Normale Superiore, I-56127 Pisa, Italy}
\author{G.~Chlachidze$^{\dag}$}~\affiliation{Fermi National Accelerator Laboratory, Batavia, Illinois 60510, USA}
\author{F.~Chlebana$^{\dag}$}~\affiliation{Fermi National Accelerator Laboratory, Batavia, Illinois 60510, USA}
\author{D.K.~Cho$^{\ddag}$}~\affiliation{Brown University, Providence, Rhode Island 02912, USA}
\author{K.~Cho$^{\dag}$}~\affiliation{Center for High Energy Physics: Kyungpook National University, Daegu 702-701, Korea; Seoul National University, Seoul 151-742, Korea; Sungkyunkwan University, Suwon 440-746, Korea; Korea Institute of Science and Technology Information, Daejeon 305-806, Korea; Chonnam National University, Gwangju 500-757, Korea; Chonbuk National University, Jeonju 561-756, Korea}
\author{S.W.~Cho$^{\ddag}$}~\affiliation{Korea Detector Laboratory, Korea University, Seoul, Korea}
\author{S.~Choi$^{\ddag}$}~\affiliation{Korea Detector Laboratory, Korea University, Seoul, Korea}
\author{D.~Chokheli$^{\dag}$}~\affiliation{Joint Institute for Nuclear Research, Dubna, Russia}
\author{B.~Choudhary$^{\ddag}$}~\affiliation{Delhi University, Delhi, India}
\author{W.H.~Chung$^{\dag}$}~\affiliation{University of Wisconsin, Madison, Wisconsin 53706, USA}
\author{Y.S.~Chung$^{\dag}$}~\affiliation{University of Rochester, Rochester, New York 14627, USA}
\author{S.~Cihangir$^{\ddag}$}~\affiliation{Fermi National Accelerator Laboratory, Batavia, Illinois 60510, USA}
\author{M.A.~Ciocci$^{\dag\sharp{d}}$}~\affiliation{Istituto Nazionale di Fisica Nucleare Pisa, $^{\sharp{c}}$University of Pisa, $^{\sharp{d}}$University of Siena and $^{\sharp{e}}$Scuola Normale Superiore, I-56127 Pisa, Italy}
\author{D.~Claes$^{\ddag}$}~\affiliation{University of Nebraska, Lincoln, Nebraska 68588, USA}
\author{A.~Clark$^{\dag}$}~\affiliation{University of Geneva, CH-1211 Geneva 4, Switzerland}
\author{C.~Clarke$^{\dag}$}~\affiliation{Wayne State University, Detroit, Michigan 48201, USA}
\author{J.~Clutter$^{\ddag}$}~\affiliation{University of Kansas, Lawrence, Kansas 66045, USA}
\author{G.~Compostella$^{\dag\sharp{b}}$}~\affiliation{Istituto Nazionale di Fisica Nucleare, Sezione di Padova-Trento, $^{\sharp{b}}$University of Padova, I-35131 Padova, Italy}
\author{M.E.~Convery$^{\dag}$}~\affiliation{Fermi National Accelerator Laboratory, Batavia, Illinois 60510, USA}
\author{J.~Conway$^{\dag}$}~\affiliation{University of California, Davis, Davis, California 95616, USA}
\author{M.~Cooke$^{\ddag}$}~\affiliation{Fermi National Accelerator Laboratory, Batavia, Illinois 60510, USA}
\author{W.E.~Cooper$^{\ddag}$}~\affiliation{Fermi National Accelerator Laboratory, Batavia, Illinois 60510, USA}
\author{M.~Corbo$^{\dag}$}~\affiliation{Fermi National Accelerator Laboratory, Batavia, Illinois 60510, USA}
\author{M.~Corcoran$^{\ddag}$}~\affiliation{Rice University, Houston, Texas 77005, USA}
\author{M.~Cordelli$^{\dag}$}~\affiliation{Laboratori Nazionali di Frascati, Istituto Nazionale di Fisica Nucleare, I-00044 Frascati, Italy}
\author{F.~Couderc$^{\ddag}$}~\affiliation{CEA, Irfu, SPP, Saclay, France}
\author{M.-C.~Cousinou$^{\ddag}$}~\affiliation{CPPM, Aix-Marseille Universit\'e, CNRS/IN2P3, Marseille, France}
\author{C.A.~Cox$^{\dag}$}~\affiliation{University of California, Davis, Davis, California 95616, USA}
\author{D.J.~Cox$^{\dag}$}~\affiliation{University of California, Davis, Davis, California 95616, USA}
\author{F.~Crescioli$^{\dag\sharp{c}}$}~\affiliation{Istituto Nazionale di Fisica Nucleare Pisa, $^{\sharp{c}}$University of Pisa, $^{\sharp{d}}$University of Siena and $^{\sharp{e}}$Scuola Normale Superiore, I-56127 Pisa, Italy}
\author{A.~Croc$^{\ddag}$}~\affiliation{CEA, Irfu, SPP, Saclay, France}
\author{J.~Cuevas$^{\dag a}$}~\affiliation{Instituto de Fisica de Cantabria, CSIC-University of Cantabria, 39005 Santander, Spain}
\author{R.~Culbertson$^{\dag}$}~\affiliation{Fermi National Accelerator Laboratory, Batavia, Illinois 60510, USA}
\author{D.~Cutts$^{\ddag}$}~\affiliation{Brown University, Providence, Rhode Island 02912, USA}
\author{D.~Dagenhart$^{\dag}$}~\affiliation{Fermi National Accelerator Laboratory, Batavia, Illinois 60510, USA}
\author{N.~d'Ascenzo$^{\dag g}$}~\affiliation{Fermi National Accelerator Laboratory, Batavia, Illinois 60510, USA}
\author{A.~Das$^{\ddag}$}~\affiliation{University of Arizona, Tucson, Arizona 85721, USA}
\author{M.~Datta$^{\dag}$}~\affiliation{Fermi National Accelerator Laboratory, Batavia, Illinois 60510, USA}
\author{G.~Davies$^{\ddag}$}~\affiliation{Imperial College London, London SW7 2AZ, United Kingdom}
\author{P.~de~Barbaro$^{\dag}$}~\affiliation{University of Rochester, Rochester, New York 14627, USA}
\author{S.J.~de~Jong$^{\ddag}$}~\affiliation{Nikhef, Science Park, Amsterdam, the Netherlands}~\affiliation{Radboud University Nijmegen, Nijmegen, the Netherlands}
\author{E.~De~La~Cruz-Burelo$^{\ddag}$}~\affiliation{CINVESTAV, Mexico City, Mexico}
\author{F.~D\'eliot$^{\ddag}$}~\affiliation{CEA, Irfu, SPP, Saclay, France}
\author{M.~Dell'Orso$^{\dag\sharp{c}}$}~\affiliation{Istituto Nazionale di Fisica Nucleare Pisa, $^{\sharp{c}}$University of Pisa, $^{\sharp{d}}$University of Siena and $^{\sharp{e}}$Scuola Normale Superiore, I-56127 Pisa, Italy}
\author{R.~Demina$^{\ddag}$}~\affiliation{University of Rochester, Rochester, New York 14627, USA}
\author{L.~Demortier$^{\dag}$}~\affiliation{The Rockefeller University, New York, New York 10065, USA}
\author{M.~Deninno$^{\dag}$}~\affiliation{Istituto Nazionale di Fisica Nucleare Bologna, $^{\sharp{a}}$University of Bologna, I-40127 Bologna, Italy}
\author{D.~Denisov$^{\ddag}$}~\affiliation{Fermi National Accelerator Laboratory, Batavia, Illinois 60510, USA}
\author{S.P.~Denisov$^{\ddag}$}~\affiliation{Institute for High Energy Physics, Protvino, Russia}
\author{M.~d'Errico$^{\dag\sharp{b}}$}~\affiliation{Istituto Nazionale di Fisica Nucleare, Sezione di Padova-Trento, $^{\sharp{b}}$University of Padova, I-35131 Padova, Italy}
\author{S.~Desai$^{\ddag}$}~\affiliation{Fermi National Accelerator Laboratory, Batavia, Illinois 60510, USA}
\author{C.~Deterre$^{\ddag}$}~\affiliation{CEA, Irfu, SPP, Saclay, France}
\author{K.~DeVaughan$^{\ddag}$}~\affiliation{University of Nebraska, Lincoln, Nebraska 68588, USA}
\author{F.~Devoto$^{\dag}$}~\affiliation{Division of High Energy Physics, Department of Physics, University of Helsinki and Helsinki Institute of Physics, FIN-00014, Helsinki, Finland}
\author{A.~Di~Canto$^{\dag\sharp{c}}$}~\affiliation{Istituto Nazionale di Fisica Nucleare Pisa, $^{\sharp{c}}$University of Pisa, $^{\sharp{d}}$University of Siena and $^{\sharp{e}}$Scuola Normale Superiore, I-56127 Pisa, Italy}
\author{B.~Di~Ruzza$^{\dag}$}~\affiliation{Fermi National Accelerator Laboratory, Batavia, Illinois 60510, USA}
\author{H.T.~Diehl$^{\ddag}$}~\affiliation{Fermi National Accelerator Laboratory, Batavia, Illinois 60510, USA}
\author{M.~Diesburg$^{\ddag}$}~\affiliation{Fermi National Accelerator Laboratory, Batavia, Illinois 60510, USA}
\author{P.F.~Ding$^{\ddag}$}~\affiliation{The University of Manchester, Manchester M13 9PL, United Kingdom}
\author{J.R.~Dittmann$^{\dag}$}~\affiliation{Baylor University, Waco, Texas 76798, USA}
\author{A.~Dominguez$^{\ddag}$}~\affiliation{University of Nebraska, Lincoln, Nebraska 68588, USA}
\author{S.~Donati$^{\dag\sharp{c}}$}~\affiliation{Istituto Nazionale di Fisica Nucleare Pisa, $^{\sharp{c}}$University of Pisa, $^{\sharp{d}}$University of Siena and $^{\sharp{e}}$Scuola Normale Superiore, I-56127 Pisa, Italy}
\author{P.~Dong$^{\dag}$}~\affiliation{Fermi National Accelerator Laboratory, Batavia, Illinois 60510, USA}
\author{M.~D'Onofrio$^{\dag}$}~\affiliation{University of Liverpool, Liverpool L69 7ZE, United Kingdom}
\author{M.~Dorigo$^{\dag}$}~\affiliation{Istituto Nazionale di Fisica Nucleare Trieste/Udine, I-34100 Trieste, $^{\sharp{g}}$University of Udine, I-33100 Udine, Italy}
\author{T.~Dorigo$^{\dag}$}~\affiliation{Istituto Nazionale di Fisica Nucleare, Sezione di Padova-Trento, $^{\sharp{b}}$University of Padova, I-35131 Padova, Italy}
\author{A.~Dubey$^{\ddag}$}~\affiliation{Delhi University, Delhi, India}
\author{L.V.~Dudko$^{\ddag}$}~\affiliation{Moscow State University, Moscow, Russia}
\author{D.~Duggan$^{\ddag}$}~\affiliation{Rutgers University, Piscataway, New Jersey 08855, USA}
\author{A.~Duperrin$^{\ddag}$}~\affiliation{CPPM, Aix-Marseille Universit\'e, CNRS/IN2P3, Marseille, France}
\author{S.~Dutt$^{\ddag}$}~\affiliation{Panjab University, Chandigarh, India}
\author{A.~Dyshkant$^{\ddag}$}~\affiliation{Northern Illinois University, DeKalb, Illinois 60115, USA}
\author{M.~Eads$^{\ddag}$}~\affiliation{University of Nebraska, Lincoln, Nebraska 68588, USA}
\author{K.~Ebina$^{\dag}$}~\affiliation{Waseda University, Tokyo 169, Japan}
\author{D.~Edmunds$^{\ddag}$}~\affiliation{Michigan State University, East Lansing, Michigan 48824, USA}
\author{A.~Elagin$^{\dag}$}~\affiliation{Texas A\&M University, College Station, Texas 77843, USA}
\author{J.~Ellison$^{\ddag}$}~\affiliation{University of California Riverside, Riverside, California 92521, USA}
\author{V.D.~Elvira$^{\ddag}$}~\affiliation{Fermi National Accelerator Laboratory, Batavia, Illinois 60510, USA}
\author{Y.~Enari$^{\ddag}$}~\affiliation{LPNHE, Universit\'es Paris VI and VII, CNRS/IN2P3, Paris, France}
\author{A.~Eppig$^{\dag}$}~\affiliation{University of Michigan, Ann Arbor, Michigan 48109, USA}
\author{R.~Erbacher$^{\dag}$}~\affiliation{University of California, Davis, Davis, California 95616, USA}
\author{S.~Errede$^{\dag}$}~\affiliation{University of Illinois, Urbana, Illinois 61801, USA}
\author{N.~Ershaidat$^{\dag h}$}~\affiliation{Fermi National Accelerator Laboratory, Batavia, Illinois 60510, USA}
\author{R.~Eusebi$^{\dag}$}~\affiliation{Texas A\&M University, College Station, Texas 77843, USA}
\author{H.~Evans$^{\ddag}$}~\affiliation{Indiana University, Bloomington, Indiana 47405, USA}
\author{A.~Evdokimov$^{\ddag}$}~\affiliation{Brookhaven National Laboratory, Upton, New York 11973, USA}
\author{V.N.~Evdokimov$^{\ddag}$}~\affiliation{Institute for High Energy Physics, Protvino, Russia}
\author{G.~Facini$^{\ddag}$}~\affiliation{Northeastern University, Boston, Massachusetts 02115, USA}
\author{S.~Farrington$^{\dag}$}~\affiliation{University of Oxford, Oxford OX1 3RH, United Kingdom}
\author{M.~Feindt$^{\dag}$}~\affiliation{Institut f\"{u}r Experimentelle Kernphysik, Karlsruhe Institute of Technology, D-76131 Karlsruhe, Germany}
\author{L.~Feng$^{\ddag}$}~\affiliation{Northern Illinois University, DeKalb, Illinois 60115, USA}
\author{T.~Ferbel$^{\ddag}$}~\affiliation{University of Rochester, Rochester, New York 14627, USA}
\author{J.P.~Fernandez$^{\dag}$}~\affiliation{Centro de Investigaciones Energeticas Medioambientales y Tecnologicas, E-28040 Madrid, Spain}
\author{C.~Ferrazza$^{\dag}$}~\affiliation{Istituto Nazionale di Fisica Nucleare, Sezione di Roma 1, $^{\sharp{f}}$Sapienza Universit\`{a} di Roma, I-00185 Roma, Italy} 
\author{F.~Fiedler$^{\ddag}$}~\affiliation{Institut f\"ur Physik, Universit\"at Mainz, Mainz, Germany}
\author{R.~Field$^{\dag}$}~\affiliation{University of Florida, Gainesville, Florida 32611, USA}
\author{F.~Filthaut$^{\ddag}$}~\affiliation{Nikhef, Science Park, Amsterdam, the Netherlands}~\affiliation{Radboud University Nijmegen, Nijmegen, the Netherlands}
\author{W.~Fisher$^{\ddag}$}~\affiliation{Michigan State University, East Lansing, Michigan 48824, USA}
\author{H.E.~Fisk$^{\ddag}$}~\affiliation{Fermi National Accelerator Laboratory, Batavia, Illinois 60510, USA}
\author{G.~Flanagan$^{\dag i}$}~\affiliation{Fermi National Accelerator Laboratory, Batavia, Illinois 60510, USA}
\author{R.~Forrest$^{\dag}$}~\affiliation{University of California, Davis, Davis, California 95616, USA}
\author{M.~Fortner$^{\ddag}$}~\affiliation{Northern Illinois University, DeKalb, Illinois 60115, USA}
\author{H.~Fox$^{\ddag}$}~\affiliation{Lancaster University, Lancaster LA1 4YB, United Kingdom}
\author{M.J.~Frank$^{\dag}$}~\affiliation{Baylor University, Waco, Texas 76798, USA}
\author{M.~Franklin$^{\dag}$}~\affiliation{Harvard University, Cambridge, Massachusetts 02138, USA}
\author{J.C.~Freeman$^{\dag}$}~\affiliation{Fermi National Accelerator Laboratory, Batavia, Illinois 60510, USA}
\author{S.~Fuess$^{\ddag}$}~\affiliation{Fermi National Accelerator Laboratory, Batavia, Illinois 60510, USA}
\author{Y.~Funakoshi$^{\dag}$}~\affiliation{Waseda University, Tokyo 169, Japan}
\author{M.~Gallinaro$^{\dag}$}~\affiliation{The Rockefeller University, New York, New York 10065, USA}
\author{A.~Garcia-Bellido$^{\ddag}$}~\affiliation{University of Rochester, Rochester, New York 14627, USA}
\author{J.E.~Garcia$^{\dag}$}~\affiliation{University of Geneva, CH-1211 Geneva 4, Switzerland}
\author{J.A.~Garc\'{\i}a-Gonz\'alez$^{\ddag}$}~\affiliation{CINVESTAV, Mexico City, Mexico}
\author{G.A.~Garc\'ia-Guerra$^{\ddag c}$}~\affiliation{CINVESTAV, Mexico City, Mexico}
\author{A.F.~Garfinkel$^{\dag}$}~\affiliation{Purdue University, West Lafayette, Indiana 47907, USA}
\author{P.~Garosi$^{\dag\sharp{d}}$}~\affiliation{Istituto Nazionale di Fisica Nucleare Pisa, $^{\sharp{c}}$University of Pisa, $^{\sharp{d}}$University of Siena and $^{\sharp{e}}$Scuola Normale Superiore, I-56127 Pisa, Italy}
\author{V.~Gavrilov$^{\ddag}$}~\affiliation{Institution for Theoretical and Experimental Physics, ITEP, Moscow 117259, Russia}
\author{P.~Gay$^{\ddag}$}~\affiliation{LPC, Universit\'e Blaise Pascal, CNRS/IN2P3, Clermont, France}
\author{W.~Geng$^{\ddag}$}~\affiliation{CPPM, Aix-Marseille Universit\'e, CNRS/IN2P3, Marseille, France}~\affiliation{Michigan State University, East Lansing, Michigan 48824, USA}
\author{D.~Gerbaudo$^{\ddag}$}~\affiliation{Princeton University, Princeton, New Jersey 08544, USA}
\author{C.E.~Gerber$^{\ddag}$}~\affiliation{University of Illinois at Chicago, Chicago, Illinois 60607, USA}
\author{H.~Gerberich$^{\dag}$}~\affiliation{University of Illinois, Urbana, Illinois 61801, USA}
\author{E.~Gerchtein$^{\dag}$}~\affiliation{Fermi National Accelerator Laboratory, Batavia, Illinois 60510, USA}
\author{Y.~Gershtein$^{\ddag}$}~\affiliation{Rutgers University, Piscataway, New Jersey 08855, USA}
\author{S.~Giagu$^{\dag}$}~\affiliation{Istituto Nazionale di Fisica Nucleare, Sezione di Roma 1, $^{\sharp{f}}$Sapienza Universit\`{a} di Roma, I-00185 Roma, Italy}
\author{V.~Giakoumopoulou$^{\dag}$}~\affiliation{University of Athens, 157 71 Athens, Greece}
\author{P.~Giannetti$^{\dag}$}~\affiliation{Istituto Nazionale di Fisica Nucleare Pisa, $^{\sharp{c}}$University of Pisa, $^{\sharp{d}}$University of Siena and $^{\sharp{e}}$Scuola Normale Superiore, I-56127 Pisa, Italy}
\author{K.~Gibson$^{\dag}$}~\affiliation{University of Pittsburgh, Pittsburgh, Pennsylvania 15260, USA}
\author{C.M.~Ginsburg$^{\dag}$}~\affiliation{Fermi National Accelerator Laboratory, Batavia, Illinois 60510, USA}
\author{G.~Ginther$^{\ddag}$}~\affiliation{Fermi National Accelerator Laboratory, Batavia, Illinois 60510, USA}~\affiliation{University of Rochester, Rochester, New York 14627, USA}
\author{N.~Giokaris$^{\dag}$}~\affiliation{University of Athens, 157 71 Athens, Greece}
\author{P.~Giromini$^{\dag}$}~\affiliation{Laboratori Nazionali di Frascati, Istituto Nazionale di Fisica Nucleare, I-00044 Frascati, Italy}
\author{G.~Giurgiu$^{\dag}$}~\affiliation{The Johns Hopkins University, Baltimore, Maryland 21218, USA}
\author{V.~Glagolev$^{\dag}$}~\affiliation{Joint Institute for Nuclear Research, Dubna, Russia}
\author{D.~Glenzinski$^{\dag}$}~\affiliation{Fermi National Accelerator Laboratory, Batavia, Illinois 60510, USA}
\author{M.~Gold$^{\dag}$}~\affiliation{University of New Mexico, Albuquerque, New Mexico 87131, USA}
\author{D.~Goldin$^{\dag}$}~\affiliation{Texas A\&M University, College Station, Texas 77843, USA}
\author{N.~Goldschmidt$^{\dag}$}~\affiliation{University of Florida, Gainesville, Florida 32611, USA}
\author{A.~Golossanov$^{\dag}$}~\affiliation{Fermi National Accelerator Laboratory, Batavia, Illinois 60510, USA}
\author{G.~Golovanov$^{\ddag}$}~\affiliation{Joint Institute for Nuclear Research, Dubna, Russia}
\author{G.~Gomez-Ceballos$^{\dag}$}~\affiliation{Massachusetts Institute of Technology, Cambridge, Massachusetts 02139, USA}
\author{G.~Gomez$^{\dag}$}~\affiliation{Instituto de Fisica de Cantabria, CSIC-University of Cantabria, 39005 Santander, Spain}
\author{M.~Goncharov$^{\dag}$}~\affiliation{Massachusetts Institute of Technology, Cambridge, Massachusetts 02139, USA}
\author{O.~Gonz\'{a}lez$^{\dag}$}~\affiliation{Centro de Investigaciones Energeticas Medioambientales y Tecnologicas, E-28040 Madrid, Spain}
\author{I.~Gorelov$^{\dag}$}~\affiliation{University of New Mexico, Albuquerque, New Mexico 87131, USA}
\author{A.T.~Goshaw$^{\dag}$}~\affiliation{Duke University, Durham, North Carolina 27708, USA}
\author{K.~Goulianos$^{\dag}$}~\affiliation{The Rockefeller University, New York, New York 10065, USA}
\author{A.~Goussiou$^{\ddag}$}~\affiliation{University of Washington, Seattle, Washington 98195, USA}
\author{P.D.~Grannis$^{\ddag}$}~\affiliation{State University of New York, Stony Brook, New York 11794, USA}
\author{S.~Greder$^{\ddag}$}~\affiliation{IPHC, Universit\'e de Strasbourg, CNRS/IN2P3, Strasbourg, France}
\author{H.~Greenlee$^{\ddag}$}~\affiliation{Fermi National Accelerator Laboratory, Batavia, Illinois 60510, USA}
\author{G.~Grenier$^{\ddag}$}~\affiliation{IPNL, Universit\'e Lyon 1, CNRS/IN2P3, Villeurbanne, France and Universit\'e de Lyon, Lyon, France}
\author{S.~Grinstein$^{\dag}$}~\affiliation{Institut de Fisica d'Altes Energies, ICREA, Universitat Autonoma de Barcelona, E-08193, Bellaterra (Barcelona), Spain}
\author{Ph.~Gris$^{\ddag}$}~\affiliation{LPC, Universit\'e Blaise Pascal, CNRS/IN2P3, Clermont, France}
\author{J.-F.~Grivaz$^{\ddag}$}~\affiliation{LAL, Universit\'e Paris-Sud, CNRS/IN2P3, Orsay, France}
\author{A.~Grohsjean$^{\ddag d}$}~\affiliation{CEA, Irfu, SPP, Saclay, France}
\author{C.~Grosso-Pilcher$^{\dag}$}~\affiliation{Enrico Fermi Institute, University of Chicago, Chicago, Illinois 60637, USA}
\author{R.C.~Group$^{\dag}$}~\affiliation{University of Virginia, Charlottesville, Virginia 22904, USA}~\affiliation{Fermi National Accelerator Laboratory, Batavia, Illinois 60510, USA}
\author{S.~Gr\"unendahl$^{\ddag}$}~\affiliation{Fermi National Accelerator Laboratory, Batavia, Illinois 60510, USA}
\author{M.W.~Gr{\"u}newald$^{\ddag}$}~\affiliation{University College Dublin, Dublin, Ireland}
\author{T.~Guillemin$^{\ddag}$}~\affiliation{LAL, Universit\'e Paris-Sud, CNRS/IN2P3, Orsay, France}
\author{J.~Guimaraes~da~Costa$^{\dag}$}~\affiliation{Harvard University, Cambridge, Massachusetts 02138, USA}
\author{G.~Gutierrez$^{\ddag}$}~\affiliation{Fermi National Accelerator Laboratory, Batavia, Illinois 60510, USA}
\author{P.~Gutierrez$^{\ddag}$}~\affiliation{University of Oklahoma, Norman, Oklahoma 73019, USA}
\author{S.~Hagopian$^{\ddag}$}~\affiliation{Florida State University, Tallahassee, Florida 32306, USA}
\author{S.R.~Hahn$^{\dag}$}~\affiliation{Fermi National Accelerator Laboratory, Batavia, Illinois 60510, USA}
\author{J.~Haley$^{\ddag}$}~\affiliation{Northeastern University, Boston, Massachusetts 02115, USA}
\author{E.~Halkiadakis$^{\dag}$}~\affiliation{Rutgers University, Piscataway, New Jersey 08855, USA}
\author{A.~Hamaguchi$^{\dag}$}~\affiliation{Osaka City University, Osaka 588, Japan}
\author{J.Y.~Han$^{\dag}$}~\affiliation{University of Rochester, Rochester, New York 14627, USA}
\author{L.~Han$^{\ddag}$}~\affiliation{University of Science and Technology of China, Hefei, People's Republic of China}
\author{F.~Happacher$^{\dag}$}~\affiliation{Laboratori Nazionali di Frascati, Istituto Nazionale di Fisica Nucleare, I-00044 Frascati, Italy}
\author{K.~Hara$^{\dag}$}~\affiliation{University of Tsukuba, Tsukuba, Ibaraki 305, Japan}
\author{K.~Harder$^{\ddag}$}~\affiliation{The University of Manchester, Manchester M13 9PL, United Kingdom}
\author{D.~Hare$^{\dag}$}~\affiliation{Rutgers University, Piscataway, New Jersey 08855, USA}
\author{M.~Hare$^{\dag}$}~\affiliation{Tufts University, Medford, Massachusetts 02155, USA}
\author{A.~Harel$^{\ddag}$}~\affiliation{University of Rochester, Rochester, New York 14627, USA}
\author{R.F.~Harr$^{\dag}$}~\affiliation{Wayne State University, Detroit, Michigan 48201, USA}
\author{K.~Hatakeyama$^{\dag}$}~\affiliation{Baylor University, Waco, Texas 76798, USA}
\author{J.M.~Hauptman$^{\ddag}$}~\affiliation{Iowa State University, Ames, Iowa 50011, USA}
\author{C.~Hays$^{\dag}$}~\affiliation{University of Oxford, Oxford OX1 3RH, United Kingdom}
\author{J.~Hays$^{\ddag}$}~\affiliation{Imperial College London, London SW7 2AZ, United Kingdom}
\author{T.~Head$^{\ddag}$}~\affiliation{The University of Manchester, Manchester M13 9PL, United Kingdom}
\author{T.~Hebbeker$^{\ddag}$}~\affiliation{III. Physikalisches Institut A, RWTH Aachen University, Aachen, Germany}
\author{M.~Heck$^{\dag}$}~\affiliation{Institut f\"{u}r Experimentelle Kernphysik, Karlsruhe Institute of Technology, D-76131 Karlsruhe, Germany}
\author{D.~Hedin$^{\ddag}$}~\affiliation{Northern Illinois University, DeKalb, Illinois 60115, USA}
\author{H.~Hegab$^{\ddag}$}~\affiliation{Oklahoma State University, Stillwater, Oklahoma 74078, USA}
\author{J.~Heinrich$^{\dag}$}~\affiliation{University of Pennsylvania, Philadelphia, Pennsylvania 19104, USA}
\author{A.P.~Heinson$^{\ddag}$}~\affiliation{University of California Riverside, Riverside, California 92521, USA}
\author{U.~Heintz$^{\ddag}$}~\affiliation{Brown University, Providence, Rhode Island 02912, USA}
\author{C.~Hensel$^{\ddag}$}~\affiliation{II. Physikalisches Institut, Georg-August-Universit\"at G\"ottingen, G\"ottingen, Germany}
\author{I.~Heredia-De~La~Cruz$^{\ddag}$}~\affiliation{CINVESTAV, Mexico City, Mexico}
\author{M.~Herndon$^{\dag}$}~\affiliation{University of Wisconsin, Madison, Wisconsin 53706, USA}
\author{K.~Herner$^{\ddag}$}~\affiliation{University of Michigan, Ann Arbor, Michigan 48109, USA}
\author{G.~Hesketh$^{\ddag e}$}~\affiliation{The University of Manchester, Manchester M13 9PL, United Kingdom}
\author{S.~Hewamanage$^{\dag}$}~\affiliation{Baylor University, Waco, Texas 76798, USA}
\author{M.D.~Hildreth$^{\ddag}$}~\affiliation{University of Notre Dame, Notre Dame, Indiana 46556, USA}
\author{R.~Hirosky$^{\ddag}$}~\affiliation{University of Virginia, Charlottesville, Virginia 22904, USA}
\author{T.~Hoang$^{\ddag}$}~\affiliation{Florida State University, Tallahassee, Florida 32306, USA}
\author{J.D.~Hobbs$^{\ddag}$}~\affiliation{State University of New York, Stony Brook, New York 11794, USA}
\author{A.~Hocker$^{\dag}$}~\affiliation{Fermi National Accelerator Laboratory, Batavia, Illinois 60510, USA}
\author{B.~Hoeneisen$^{\ddag}$}~\affiliation{Universidad San Francisco de Quito, Quito, Ecuador}
\author{J.~Hogan$^{\ddag}$}~\affiliation{Rice University, Houston, Texas 77005, USA}
\author{M.~Hohlfeld$^{\ddag}$}~\affiliation{Institut f\"ur Physik, Universit\"at Mainz, Mainz, Germany}
\author{W.~Hopkins$^{\dag j}$}~\affiliation{Fermi National Accelerator Laboratory, Batavia, Illinois 60510, USA}
\author{D.~Horn$^{\dag}$}~\affiliation{Institut f\"{u}r Experimentelle Kernphysik, Karlsruhe Institute of Technology, D-76131 Karlsruhe, Germany}
\author{S.~Hou$^{\dag}$}~\affiliation{Institute of Physics, Academia Sinica, Taipei, Taiwan 11529, Republic of China}
\author{I.~Howley$^{\ddag}$}~\affiliation{University of Texas, Arlington, Texas 76019, USA}
\author{Z.~Hubacek$^{\ddag}$}~\affiliation{Czech Technical University in Prague, Prague, Czech Republic}~\affiliation{CEA, Irfu, SPP, Saclay, France}
\author{R.E.~Hughes$^{\dag}$}~\affiliation{The Ohio State University, Columbus, Ohio 43210, USA}
\author{M.~Hurwitz$^{\dag}$}~\affiliation{Enrico Fermi Institute, University of Chicago, Chicago, Illinois 60637, USA}
\author{U.~Husemann$^{\dag}$}~\affiliation{Yale University, New Haven, Connecticut 06520, USA}
\author{N.~Hussain$^{\dag}$}~\affiliation{Institute of Particle Physics: McGill University, Montr\'{e}al, Qu\'{e}bec, Canada H3A~2T8; Simon Fraser University, Burnaby, British Columbia, Canada V5A~1S6; University of Toronto, Toronto, Ontario, Canada M5S~1A7; and TRIUMF, Vancouver, British Columbia, Canada V6T~2A3}
\author{M.~Hussein$^{\dag}$}~\affiliation{Michigan State University, East Lansing, Michigan 48824, USA}
\author{J.~Huston$^{\dag}$}~\affiliation{Michigan State University, East Lansing, Michigan 48824, USA}
\author{V.~Hynek$^{\ddag}$}~\affiliation{Czech Technical University in Prague, Prague, Czech Republic}
\author{I.~Iashvili$^{\ddag}$}~\affiliation{State University of New York, Buffalo, New York 14260, USA}
\author{Y.~Ilchenko$^{\ddag}$}~\affiliation{Southern Methodist University, Dallas, Texas 75275, USA}
\author{R.~Illingworth$^{\ddag}$}~\affiliation{Fermi National Accelerator Laboratory, Batavia, Illinois 60510, USA}
\author{G.~Introzzi$^{\dag}$}~\affiliation{Istituto Nazionale di Fisica Nucleare Pisa, $^{\sharp{c}}$University of Pisa, $^{\sharp{d}}$University of Siena and $^{\sharp{e}}$Scuola Normale Superiore, I-56127 Pisa, Italy}
\author{M.~Iori$^{\dag\sharp{f}}$}~\affiliation{Istituto Nazionale di Fisica Nucleare, Sezione di Roma 1, $^{\sharp{f}}$Sapienza Universit\`{a} di Roma, I-00185 Roma, Italy}
\author{A.S.~Ito$^{\ddag}$}~\affiliation{Fermi National Accelerator Laboratory, Batavia, Illinois 60510, USA}
\author{A.~Ivanov$^{\dag k}$}~\affiliation{University of California, Davis, Davis, California 95616, USA}
\author{S.~Jabeen$^{\ddag}$}~\affiliation{Brown University, Providence, Rhode Island 02912, USA}
\author{M.~Jaffr\'e$^{\ddag}$}~\affiliation{LAL, Universit\'e Paris-Sud, CNRS/IN2P3, Orsay, France}
\author{E.~James$^{\dag}$}~\affiliation{Fermi National Accelerator Laboratory, Batavia, Illinois 60510, USA}
\author{D.~Jang$^{\dag}$}~\affiliation{Carnegie Mellon University, Pittsburgh, Pennsylvania 15213, USA}
\author{A.~Jayasinghe$^{\ddag}$}~\affiliation{University of Oklahoma, Norman, Oklahoma 73019, USA}
\author{B.~Jayatilaka$^{\dag}$}~\affiliation{Duke University, Durham, North Carolina 27708, USA}
\author{D.T.~Jeans$^{\dag}$}~\affiliation{Istituto Nazionale di Fisica Nucleare, Sezione di Roma 1, $^{\sharp{f}}$Sapienza Universit\`{a} di Roma, I-00185 Roma, Italy}
\author{E.J.~Jeon$^{\dag}$}~\affiliation{Center for High Energy Physics: Kyungpook National University, Daegu 702-701, Korea; Seoul National University, Seoul 151-742, Korea; Sungkyunkwan University, Suwon 440-746, Korea; Korea Institute of Science and Technology Information, Daejeon 305-806, Korea; Chonnam National University, Gwangju 500-757, Korea; Chonbuk National University, Jeonju 561-756, Korea}
\author{M.S.~Jeong$^{\ddag}$}~\affiliation{Korea Detector Laboratory, Korea University, Seoul, Korea}
\author{R.~Jesik$^{\ddag}$}~\affiliation{Imperial College London, London SW7 2AZ, United Kingdom}
\author{P.~Jiang$^{\ddag}$} \affiliation{University of Science and Technology of China, Hefei, People's Republic of China}
\author{S.~Jindariani$^{\dag}$}~\affiliation{Fermi National Accelerator Laboratory, Batavia, Illinois 60510, USA}
\author{K.~Johns$^{\ddag}$}~\affiliation{University of Arizona, Tucson, Arizona 85721, USA}
\author{E.~Johnson$^{\ddag}$}~\affiliation{Michigan State University, East Lansing, Michigan 48824, USA}
\author{M.~Johnson$^{\ddag}$}~\affiliation{Fermi National Accelerator Laboratory, Batavia, Illinois 60510, USA}
\author{A.~Jonckheere$^{\ddag}$}~\affiliation{Fermi National Accelerator Laboratory, Batavia, Illinois 60510, USA}
\author{M.~Jones$^{\dag}$}~\affiliation{Purdue University, West Lafayette, Indiana 47907, USA}
\author{P.~Jonsson$^{\ddag}$}~\affiliation{Imperial College London, London SW7 2AZ, United Kingdom}
\author{K.K.~Joo$^{\dag}$}~\affiliation{Center for High Energy Physics: Kyungpook National University, Daegu 702-701, Korea; Seoul National University, Seoul 151-742, Korea; Sungkyunkwan University, Suwon 440-746, Korea; Korea Institute of Science and Technology Information, Daejeon 305-806, Korea; Chonnam National University, Gwangju 500-757, Korea; Chonbuk National University, Jeonju 561-756, Korea}
\author{J.~Joshi$^{\ddag}$}~\affiliation{University of California Riverside, Riverside, California 92521, USA}
\author{S.Y.~Jun$^{\dag}$}~\affiliation{Carnegie Mellon University, Pittsburgh, Pennsylvania 15213, USA}
\author{A.W.~Jung$^{\ddag}$}~\affiliation{Fermi National Accelerator Laboratory, Batavia, Illinois 60510, USA}
\author{T.R.~Junk$^{\dag}$}~\affiliation{Fermi National Accelerator Laboratory, Batavia, Illinois 60510, USA}
\author{A.~Juste$^{\ddag}$}~\affiliation{Instituci\'{o} Catalana de Recerca i Estudis Avan\c{c}ats (ICREA) and Institut de F\'{i}sica d'Altes Energies (IFAE), Barcelona, Spain}
\author{K.~Kaadze$^{\ddag}$}~\affiliation{Kansas State University, Manhattan, Kansas 66506, USA}
\author{E.~Kajfasz$^{\ddag}$}~\affiliation{CPPM, Aix-Marseille Universit\'e, CNRS/IN2P3, Marseille, France}
\author{T.~Kamon$^{\dag}$}~\affiliation{Texas A\&M University, College Station, Texas 77843, USA}
\author{P.E.~Karchin$^{\dag}$}~\affiliation{Wayne State University, Detroit, Michigan 48201, USA}
\author{D.~Karmanov$^{\ddag}$}~\affiliation{Moscow State University, Moscow, Russia}
\author{A.~Kasmi$^{\dag}$}~\affiliation{Baylor University, Waco, Texas 76798, USA}
\author{P.A.~Kasper$^{\ddag}$}~\affiliation{Fermi National Accelerator Laboratory, Batavia, Illinois 60510, USA}
\author{Y.~Kato$^{\dag l}$}~\affiliation{Osaka City University, Osaka 588, Japan}
\author{I.~Katsanos$^{\ddag}$}~\affiliation{University of Nebraska, Lincoln, Nebraska 68588, USA}
\author{R.~Kehoe$^{\ddag}$}~\affiliation{Southern Methodist University, Dallas, Texas 75275, USA}
\author{S.~Kermiche$^{\ddag}$}~\affiliation{CPPM, Aix-Marseille Universit\'e, CNRS/IN2P3, Marseille, France}
\author{W.~Ketchum$^{\dag}$}~\affiliation{Enrico Fermi Institute, University of Chicago, Chicago, Illinois 60637, USA}
\author{J.~Keung$^{\dag}$}~\affiliation{University of Pennsylvania, Philadelphia, Pennsylvania 19104, USA}
\author{N.~Khalatyan$^{\ddag}$}~\affiliation{Fermi National Accelerator Laboratory, Batavia, Illinois 60510, USA}
\author{A.~Khanov$^{\ddag}$}~\affiliation{Oklahoma State University, Stillwater, Oklahoma 74078, USA}
\author{A.~Kharchilava$^{\ddag}$}~\affiliation{State University of New York, Buffalo, New York 14260, USA}
\author{Y.N.~Kharzheev$^{\ddag}$}~\affiliation{Joint Institute for Nuclear Research, Dubna, Russia}
\author{V.~Khotilovich$^{\dag}$}~\affiliation{Texas A\&M University, College Station, Texas 77843, USA}
\author{B.~Kilminster$^{\dag}$}~\affiliation{Fermi National Accelerator Laboratory, Batavia, Illinois 60510, USA}
\author{D.H.~Kim$^{\dag}$}~\affiliation{Center for High Energy Physics: Kyungpook National University, Daegu 702-701, Korea; Seoul National University, Seoul 151-742, Korea; Sungkyunkwan University, Suwon 440-746, Korea; Korea Institute of Science and Technology Information, Daejeon 305-806, Korea; Chonnam National University, Gwangju 500-757, Korea; Chonbuk National University, Jeonju 561-756, Korea}
\author{H.S.~Kim$^{\dag}$}~\affiliation{Center for High Energy Physics: Kyungpook National University, Daegu 702-701, Korea; Seoul National University, Seoul 151-742, Korea; Sungkyunkwan University, Suwon 440-746, Korea; Korea Institute of Science and Technology Information, Daejeon 305-806, Korea; Chonnam National University, Gwangju 500-757, Korea; Chonbuk National University, Jeonju 561-756, Korea}
\author{J.E.~Kim$^{\dag}$}~\affiliation{Center for High Energy Physics: Kyungpook National University, Daegu 702-701, Korea; Seoul National University, Seoul 151-742, Korea; Sungkyunkwan University, Suwon 440-746, Korea; Korea Institute of Science and Technology Information, Daejeon 305-806, Korea; Chonnam National University, Gwangju 500-757, Korea; Chonbuk National University, Jeonju 561-756, Korea}
\author{M.J.~Kim$^{\dag}$}~\affiliation{Laboratori Nazionali di Frascati, Istituto Nazionale di Fisica Nucleare, I-00044 Frascati, Italy}
\author{S.B.~Kim$^{\dag}$}~\affiliation{Center for High Energy Physics: Kyungpook National University, Daegu 702-701, Korea; Seoul National University, Seoul 151-742, Korea; Sungkyunkwan University, Suwon 440-746, Korea; Korea Institute of Science and Technology Information, Daejeon 305-806, Korea; Chonnam National University, Gwangju 500-757, Korea; Chonbuk National University, Jeonju 561-756, Korea}
\author{S.H.~Kim$^{\dag}$}~\affiliation{University of Tsukuba, Tsukuba, Ibaraki 305, Japan}
\author{Y.J.~Kim$^{\dag}$}~\affiliation{Center for High Energy Physics: Kyungpook National University, Daegu 702-701, Korea; Seoul National University, Seoul 151-742, Korea; Sungkyunkwan University, Suwon 440-746, Korea; Korea Institute of Science and Technology Information, Daejeon 305-806, Korea; Chonnam National University, Gwangju 500-757, Korea; Chonbuk National University, Jeonju 561-756, Korea}
\author{Y.K.~Kim$^{\dag}$}~\affiliation{Enrico Fermi Institute, University of Chicago, Chicago, Illinois 60637, USA}
\author{N.~Kimura$^{\dag}$}~\affiliation{Waseda University, Tokyo 169, Japan}
\author{M.~Kirby$^{\dag}$}~\affiliation{Fermi National Accelerator Laboratory, Batavia, Illinois 60510, USA}
\author{I.~Kiselevich$^{\ddag}$}~\affiliation{Institution for Theoretical and Experimental Physics, ITEP, Moscow 117259, Russia}
\author{S.~Klimenko$^{\dag}$}~\affiliation{University of Florida, Gainesville, Florida 32611, USA}
\author{K.~Knoepfel$^{\dag}$}~\affiliation{Fermi National Accelerator Laboratory, Batavia, Illinois 60510, USA}
\author{J.M.~Kohli$^{\ddag}$}~\affiliation{Panjab University, Chandigarh, India}
\author{K.~Kondo\footnote{Deceased}$^{\dag}$}~\affiliation{Waseda University, Tokyo 169, Japan}
\author{D.J.~Kong$^{\dag}$}~\affiliation{Center for High Energy Physics: Kyungpook National University, Daegu 702-701, Korea; Seoul National University, Seoul 151-742, Korea; Sungkyunkwan University, Suwon 440-746, Korea; Korea Institute of Science and Technology Information, Daejeon 305-806, Korea; Chonnam National University, Gwangju 500-757, Korea; Chonbuk National University, Jeonju 561-756, Korea}
\author{J.~Konigsberg$^{\dag}$}~\affiliation{University of Florida, Gainesville, Florida 32611, USA}
\author{A.V.~Kotwal$^{\dag}$}~\affiliation{Duke University, Durham, North Carolina 27708, USA}
\author{A.V.~Kozelov$^{\ddag}$}~\affiliation{Institute for High Energy Physics, Protvino, Russia}
\author{J.~Kraus$^{\ddag}$}~\affiliation{University of Mississippi, University, Mississippi 38677, USA}
\author{M.~Kreps$^{\dag}$}~\affiliation{Institut f\"{u}r Experimentelle Kernphysik, Karlsruhe Institute of Technology, D-76131 Karlsruhe, Germany}
\author{J.~Kroll$^{\dag}$}~\affiliation{University of Pennsylvania, Philadelphia, Pennsylvania 19104, USA}
\author{D.~Krop$^{\dag}$}~\affiliation{Enrico Fermi Institute, University of Chicago, Chicago, Illinois 60637, USA}
\author{M.~Kruse$^{\dag}$}~\affiliation{Duke University, Durham, North Carolina 27708, USA}
\author{V.~Krutelyov$^{\dag m}$}~\affiliation{Texas A\&M University, College Station, Texas 77843, USA}
\author{T.~Kuhr$^{\dag}$}~\affiliation{Institut f\"{u}r Experimentelle Kernphysik, Karlsruhe Institute of Technology, D-76131 Karlsruhe, Germany}
\author{S.~Kulikov$^{\ddag}$}~\affiliation{Institute for High Energy Physics, Protvino, Russia}
\author{A.~Kumar$^{\ddag}$}~\affiliation{State University of New York, Buffalo, New York 14260, USA}
\author{A.~Kupco$^{\ddag}$}~\affiliation{Center for Particle Physics, Institute of Physics, Academy of Sciences of the Czech Republic, Prague, Czech Republic}
\author{M.~Kurata$^{\dag}$}~\affiliation{University of Tsukuba, Tsukuba, Ibaraki 305, Japan}
\author{T.~Kur\v{c}a$^{\ddag}$}~\affiliation{IPNL, Universit\'e Lyon 1, CNRS/IN2P3, Villeurbanne, France and Universit\'e de Lyon, Lyon, France}
\author{V.A.~Kuzmin$^{\ddag}$}~\affiliation{Moscow State University, Moscow, Russia}
\author{S.~Kwang$^{\dag}$}~\affiliation{Enrico Fermi Institute, University of Chicago, Chicago, Illinois 60637, USA}
\author{A.T.~Laasanen$^{\dag}$}~\affiliation{Purdue University, West Lafayette, Indiana 47907, USA}
\author{S.~Lami$^{\dag}$}~\affiliation{Istituto Nazionale di Fisica Nucleare Pisa, $^{\sharp{c}}$University of Pisa, $^{\sharp{d}}$University of Siena and $^{\sharp{e}}$Scuola Normale Superiore, I-56127 Pisa, Italy}
\author{S.~Lammel$^{\dag}$}~\affiliation{Fermi National Accelerator Laboratory, Batavia, Illinois 60510, USA}
\author{S.~Lammers$^{\ddag}$}~\affiliation{Indiana University, Bloomington, Indiana 47405, USA}
\author{M.~Lancaster$^{\dag}$}~\affiliation{University College London, London WC1E 6BT, United Kingdom}
\author{R.L.~Lander$^{\dag}$}~\affiliation{University of California, Davis, Davis, California 95616, USA}
\author{K.~Lannon$^{\dag n}$}~\affiliation{The Ohio State University, Columbus, Ohio 43210, USA}
\author{A.~Lath$^{\dag}$}~\affiliation{Rutgers University, Piscataway, New Jersey 08855, USA}
\author{G.~Latino$^{\dag\sharp{d}}$}~\affiliation{Istituto Nazionale di Fisica Nucleare Pisa, $^{\sharp{c}}$University of Pisa, $^{\sharp{d}}$University of Siena and $^{\sharp{e}}$Scuola Normale Superiore, I-56127 Pisa, Italy}
\author{P.~Lebrun$^{\ddag}$}~\affiliation{IPNL, Universit\'e Lyon 1, CNRS/IN2P3, Villeurbanne, France and Universit\'e de Lyon, Lyon, France}
\author{T.~LeCompte$^{\dag}$}~\affiliation{Argonne National Laboratory, Argonne, Illinois 60439, USA}
\author{E.~Lee$^{\dag}$}~\affiliation{Texas A\&M University, College Station, Texas 77843, USA}
\author{H.S.~Lee$^{\ddag}$}~\affiliation{Korea Detector Laboratory, Korea University, Seoul, Korea}
\author{H.S.~Lee$^{\dag o}$}~\affiliation{Enrico Fermi Institute, University of Chicago, Chicago, Illinois 60637, USA}
\author{J.S.~Lee$^{\dag}$}~\affiliation{Center for High Energy Physics: Kyungpook National University, Daegu 702-701, Korea; Seoul National University, Seoul 151-742, Korea; Sungkyunkwan University, Suwon 440-746, Korea; Korea Institute of Science and Technology Information, Daejeon 305-806, Korea; Chonnam National University, Gwangju 500-757, Korea; Chonbuk National University, Jeonju 561-756, Korea}
\author{S.W.~Lee$^{\dag p}$}~\affiliation{Texas A\&M University, College Station, Texas 77843, USA}
\author{S.W.~Lee$^{\ddag}$}~\affiliation{Iowa State University, Ames, Iowa 50011, USA}
\author{W.M.~Lee$^{\ddag}$}~\affiliation{Fermi National Accelerator Laboratory, Batavia, Illinois 60510, USA}
\author{X.~Lei$^{\ddag}$}~\affiliation{University of Arizona, Tucson, Arizona 85721, USA}
\author{J.~Lellouch$^{\ddag}$}~\affiliation{LPNHE, Universit\'es Paris VI and VII, CNRS/IN2P3, Paris, France}
\author{S.~Leo$^{\dag\sharp{c}}$}~\affiliation{Istituto Nazionale di Fisica Nucleare Pisa, $^{\sharp{c}}$University of Pisa, $^{\sharp{d}}$University of Siena and $^{\sharp{e}}$Scuola Normale Superiore, I-56127 Pisa, Italy}
\author{S.~Leone$^{\dag}$}~\affiliation{Istituto Nazionale di Fisica Nucleare Pisa, $^{\sharp{c}}$University of Pisa, $^{\sharp{d}}$University of Siena and $^{\sharp{e}}$Scuola Normale Superiore, I-56127 Pisa, Italy}
\author{J.D.~Lewis$^{\dag}$}~\affiliation{Fermi National Accelerator Laboratory, Batavia, Illinois 60510, USA}
\author{D.~Li$^{\ddag}$} \affiliation{LPNHE, Universit\'es Paris VI and VII, CNRS/IN2P3, Paris, France}
\author{H.~Li$^{\ddag}$}~\affiliation{LPSC, Universit\'e Joseph Fourier Grenoble 1, CNRS/IN2P3, Institut National Polytechnique de Grenoble, Grenoble, France}
\author{L.~Li$^{\ddag}$}~\affiliation{University of California Riverside, Riverside, California 92521, USA}
\author{Q.Z.~Li$^{\ddag}$}~\affiliation{Fermi National Accelerator Laboratory, Batavia, Illinois 60510, USA}
\author{J.K.~Lim$^{\ddag}$}~\affiliation{Korea Detector Laboratory, Korea University, Seoul, Korea}
\author{A.~Limosani$^{\dag q}$}~\affiliation{Duke University, Durham, North Carolina 27708, USA}
\author{D.~Lincoln$^{\ddag}$}~\affiliation{Fermi National Accelerator Laboratory, Batavia, Illinois 60510, USA}
\author{C.-J.~Lin$^{\dag}$}~\affiliation{Ernest Orlando Lawrence Berkeley National Laboratory, Berkeley, California 94720, USA}
\author{M.~Lindgren$^{\dag}$}~\affiliation{Fermi National Accelerator Laboratory, Batavia, Illinois 60510, USA}
\author{J.~Linnemann$^{\ddag}$}~\affiliation{Michigan State University, East Lansing, Michigan 48824, USA}
\author{V.V.~Lipaev$^{\ddag}$}~\affiliation{Institute for High Energy Physics, Protvino, Russia}
\author{E.~Lipeles$^{\dag}$}~\affiliation{University of Pennsylvania, Philadelphia, Pennsylvania 19104, USA}
\author{R.~Lipton$^{\ddag}$}~\affiliation{Fermi National Accelerator Laboratory, Batavia, Illinois 60510, USA}
\author{A.~Lister$^{\dag}$}~\affiliation{University of Geneva, CH-1211 Geneva 4, Switzerland}
\author{D.O.~Litvintsev$^{\dag}$}~\affiliation{Fermi National Accelerator Laboratory, Batavia, Illinois 60510, USA}
\author{C.~Liu$^{\dag}$}~\affiliation{University of Pittsburgh, Pittsburgh, Pennsylvania 15260, USA}
\author{H.~Liu$^{\dag}$}~\affiliation{University of Virginia, Charlottesville, Virginia 22904, USA}
\author{H.~Liu$^{\ddag}$}~\affiliation{Southern Methodist University, Dallas, Texas 75275, USA}
\author{Q.~Liu$^{\dag}$}~\affiliation{Purdue University, West Lafayette, Indiana 47907, USA}
\author{T.~Liu$^{\dag}$}~\affiliation{Fermi National Accelerator Laboratory, Batavia, Illinois 60510, USA}
\author{Y.~Liu$^{\ddag}$}~\affiliation{University of Science and Technology of China, Hefei, People's Republic of China}
\author{A.~Lobodenko$^{\ddag}$}~\affiliation{Petersburg Nuclear Physics Institute, St. Petersburg, Russia}
\author{S.~Lockwitz$^{\dag}$}~\affiliation{Yale University, New Haven, Connecticut 06520, USA}
\author{A.~Loginov$^{\dag}$}~\affiliation{Yale University, New Haven, Connecticut 06520, USA}
\author{M.~Lokajicek$^{\ddag}$}~\affiliation{Center for Particle Physics, Institute of Physics, Academy of Sciences of the Czech Republic, Prague, Czech Republic}
\author{R.~Lopes~de~Sa$^{\ddag}$}~\affiliation{State University of New York, Stony Brook, New York 11794, USA}
\author{H.J.~Lubatti$^{\ddag}$}~\affiliation{University of Washington, Seattle, Washington 98195, USA}
\author{D.~Lucchesi$^{\dag\sharp{b}}$}~\affiliation{Istituto Nazionale di Fisica Nucleare, Sezione di Padova-Trento, $^{\sharp{b}}$University of Padova, I-35131 Padova, Italy}
\author{J.~Lueck$^{\dag}$}~\affiliation{Institut f\"{u}r Experimentelle Kernphysik, Karlsruhe Institute of Technology, D-76131 Karlsruhe, Germany}
\author{P.~Lujan$^{\dag}$}~\affiliation{Ernest Orlando Lawrence Berkeley National Laboratory, Berkeley, California 94720, USA}
\author{P.~Lukens$^{\dag}$}~\affiliation{Fermi National Accelerator Laboratory, Batavia, Illinois 60510, USA}
\author{R.~Luna-Garcia$^{\ddag f}$}~\affiliation{CINVESTAV, Mexico City, Mexico}
\author{G.~Lungu$^{\dag}$}~\affiliation{The Rockefeller University, New York, New York 10065, USA}
\author{A.L.~Lyon$^{\ddag}$}~\affiliation{Fermi National Accelerator Laboratory, Batavia, Illinois 60510, USA}
\author{R.~Lysak$^{\dag r}$}~\affiliation{Comenius University, 842 48 Bratislava, Slovakia; Institute of Experimental Physics, 040 01 Kosice, Slovakia}
\author{J.~Lys$^{\dag}$}~\affiliation{Ernest Orlando Lawrence Berkeley National Laboratory, Berkeley, California 94720, USA}
\author{A.K.A.~Maciel$^{\ddag}$}~\affiliation{LAFEX, Centro Brasileiro de Pesquisas F\'{i}sicas, Rio de Janeiro, Brazil}
\author{R.~Madar$^{\ddag}$}~\affiliation{CEA, Irfu, SPP, Saclay, France}
\author{R.~Madrak$^{\dag}$}~\affiliation{Fermi National Accelerator Laboratory, Batavia, Illinois 60510, USA}
\author{P.~Maestro$^{\dag\sharp{d}}$}~\affiliation{Istituto Nazionale di Fisica Nucleare Pisa, $^{\sharp{c}}$University of Pisa, $^{\sharp{d}}$University of Siena and $^{\sharp{e}}$Scuola Normale Superiore, I-56127 Pisa, Italy}
\author{R.~Maga\~na-Villalba$^{\ddag}$}~\affiliation{CINVESTAV, Mexico City, Mexico}
\author{S.~Malik$^{\dag}$}~\affiliation{The Rockefeller University, New York, New York 10065, USA}
\author{S.~Malik$^{\ddag}$}~\affiliation{University of Nebraska, Lincoln, Nebraska 68588, USA}
\author{V.L.~Malyshev$^{\ddag}$}~\affiliation{Joint Institute for Nuclear Research, Dubna, Russia}
\author{G.~Manca$^{\dag s}$}~\affiliation{University of Liverpool, Liverpool L69 7ZE, United Kingdom}
\author{A.~Manousakis-Katsikakis$^{\dag}$}~\affiliation{University of Athens, 157 71 Athens, Greece}
\author{Y.~Maravin$^{\ddag}$}~\affiliation{Kansas State University, Manhattan, Kansas 66506, USA}
\author{F.~Margaroli$^{\dag}$}~\affiliation{Istituto Nazionale di Fisica Nucleare, Sezione di Roma 1, $^{\sharp{f}}$Sapienza Universit\`{a} di Roma, I-00185 Roma, Italy}
\author{C.~Marino$^{\dag}$}~\affiliation{Institut f\"{u}r Experimentelle Kernphysik, Karlsruhe Institute of Technology, D-76131 Karlsruhe, Germany}
\author{M.~Mart\'{\i}nez$^{\dag}$}~\affiliation{Institut de Fisica d'Altes Energies, ICREA, Universitat Autonoma de Barcelona, E-08193, Bellaterra (Barcelona), Spain}
\author{J.~Mart\'{\i}nez-Ortega$^{\ddag}$}~\affiliation{CINVESTAV, Mexico City, Mexico}
\author{P.~Mastrandrea$^{\dag}$}~\affiliation{Istituto Nazionale di Fisica Nucleare, Sezione di Roma 1, $^{\sharp{f}}$Sapienza Universit\`{a} di Roma, I-00185 Roma, Italy}
\author{K.~Matera$^{\dag}$}~\affiliation{University of Illinois, Urbana, Illinois 61801, USA}
\author{M.E.~Mattson$^{\dag}$}~\affiliation{Wayne State University, Detroit, Michigan 48201, USA}
\author{A.~Mazzacane$^{\dag}$}~\affiliation{Fermi National Accelerator Laboratory, Batavia, Illinois 60510, USA}
\author{P.~Mazzanti$^{\dag}$}~\affiliation{Istituto Nazionale di Fisica Nucleare Bologna, $^{\sharp{a}}$University of Bologna, I-40127 Bologna, Italy}
\author{R.~McCarthy$^{\ddag}$}~\affiliation{State University of New York, Stony Brook, New York 11794, USA}
\author{K.S.~McFarland$^{\dag}$}~\affiliation{University of Rochester, Rochester, New York 14627, USA}
\author{C.L.~McGivern$^{\ddag}$}~\affiliation{The University of Manchester, Manchester M13 9PL, United Kingdom}
\author{P.~McIntyre$^{\dag}$}~\affiliation{Texas A\&M University, College Station, Texas 77843, USA}
\author{R.~McNulty$^{\dag t}$}~\affiliation{University of Liverpool, Liverpool L69 7ZE, United Kingdom}
\author{A.~Mehta$^{\dag}$}~\affiliation{University of Liverpool, Liverpool L69 7ZE, United Kingdom}
\author{P.~Mehtala$^{\dag}$}~\affiliation{Division of High Energy Physics, Department of Physics, University of Helsinki and Helsinki Institute of Physics, FIN-00014, Helsinki, Finland}
\author{M.M.~Meijer$^{\ddag}$}~\affiliation{Nikhef, Science Park, Amsterdam, the Netherlands}~\affiliation{Radboud University Nijmegen, Nijmegen, the Netherlands}
\author{A.~Melnitchouk$^{\ddag}$}~\affiliation{University of Mississippi, University, Mississippi 38677, USA}
\author{D.~Menezes$^{\ddag}$}~\affiliation{Northern Illinois University, DeKalb, Illinois 60115, USA}
\author{P.G.~Mercadante$^{\ddag}$}~\affiliation{Universidade Federal do ABC, Santo Andr\'e, Brazil}
\author{M.~Merkin$^{\ddag}$}~\affiliation{Moscow State University, Moscow, Russia}
\author{C.~Mesropian$^{\dag}$}\affiliation{The Rockefeller University, New York, New York 10065, USA}
\author{A.~Meyer$^{\ddag}$}~\affiliation{III. Physikalisches Institut A, RWTH Aachen University, Aachen, Germany}
\author{J.~Meyer$^{\ddag}$}~\affiliation{II. Physikalisches Institut, Georg-August-Universit\"at G\"ottingen, G\"ottingen, Germany}
\author{T.~Miao$^{\dag}$}~\affiliation{Fermi National Accelerator Laboratory, Batavia, Illinois 60510, USA}
\author{F.~Miconi$^{\ddag}$}~\affiliation{IPHC, Universit\'e de Strasbourg, CNRS/IN2P3, Strasbourg, France}
\author{D.~Mietlicki$^{\dag}$}~\affiliation{University of Michigan, Ann Arbor, Michigan 48109, USA}
\author{A.~Mitra$^{\dag}$}~\affiliation{Institute of Physics, Academia Sinica, Taipei, Taiwan 11529, Republic of China}
\author{H.~Miyake$^{\dag}$}~\affiliation{University of Tsukuba, Tsukuba, Ibaraki 305, Japan}
\author{S.~Moed$^{\dag}$}~\affiliation{Fermi National Accelerator Laboratory, Batavia, Illinois 60510, USA}
\author{N.~Moggi$^{\dag}$}~\affiliation{Istituto Nazionale di Fisica Nucleare Bologna, $^{\sharp{a}}$University of Bologna, I-40127 Bologna, Italy}
\author{N.K.~Mondal$^{\ddag}$}~\affiliation{Tata Institute of Fundamental Research, Mumbai, India}
\author{M.N.~Mondragon$^{\dag c}$}~\affiliation{Fermi National Accelerator Laboratory, Batavia, Illinois 60510, USA}
\author{C.S.~Moon$^{\dag}$}~\affiliation{Center for High Energy Physics: Kyungpook National University, Daegu 702-701, Korea; Seoul National University, Seoul 151-742, Korea; Sungkyunkwan University, Suwon 440-746, Korea; Korea Institute of Science and Technology Information, Daejeon 305-806, Korea; Chonnam National University, Gwangju 500-757, Korea; Chonbuk National University, Jeonju 561-756, Korea}
\author{R.~Moore$^{\dag}$}~\affiliation{Fermi National Accelerator Laboratory, Batavia, Illinois 60510, USA}
\author{M.J.~Morello$^{\dag\sharp{e}}$}~\affiliation{Istituto Nazionale di Fisica Nucleare Pisa, $^{\sharp{c}}$University of Pisa, $^{\sharp{d}}$University of Siena and $^{\sharp{e}}$Scuola Normale Superiore, I-56127 Pisa, Italy}
\author{J.~Morlock$^{\dag}$}~\affiliation{Institut f\"{u}r Experimentelle Kernphysik, Karlsruhe Institute of Technology, D-76131 Karlsruhe, Germany}
\author{P.~Movilla~Fernandez$^{\dag}$}~\affiliation{Fermi National Accelerator Laboratory, Batavia, Illinois 60510, USA}
\author{A.~Mukherjee$^{\dag}$}~\affiliation{Fermi National Accelerator Laboratory, Batavia, Illinois 60510, USA}
\author{M.~Mulhearn$^{\ddag}$}~\affiliation{University of Virginia, Charlottesville, Virginia 22904, USA}
\author{Th.~Muller$^{\dag}$}~\affiliation{Institut f\"{u}r Experimentelle Kernphysik, Karlsruhe Institute of Technology, D-76131 Karlsruhe, Germany}
\author{P.~Murat$^{\dag}$}~\affiliation{Fermi National Accelerator Laboratory, Batavia, Illinois 60510, USA}
\author{M.~Mussini$^{\dag\sharp{a}}$}~\affiliation{Istituto Nazionale di Fisica Nucleare Bologna, $^{\sharp{a}}$University of Bologna, I-40127 Bologna, Italy}
\author{J.~Nachtman$^{\dag u}$}~\affiliation{Fermi National Accelerator Laboratory, Batavia, Illinois 60510, USA}
\author{Y.~Nagai$^{\dag}$}~\affiliation{University of Tsukuba, Tsukuba, Ibaraki 305, Japan}
\author{J.~Naganoma$^{\dag}$}~\affiliation{Waseda University, Tokyo 169, Japan}
\author{E.~Nagy$^{\ddag}$}~\affiliation{CPPM, Aix-Marseille Universit\'e, CNRS/IN2P3, Marseille, France}
\author{M.~Naimuddin$^{\ddag}$}~\affiliation{Delhi University, Delhi, India}
\author{I.~Nakano$^{\dag}$}~\affiliation{Okayama University, Okayama 700-8530, Japan}
\author{A.~Napier$^{\dag}$}~\affiliation{Tufts University, Medford, Massachusetts 02155, USA}
\author{M.~Narain$^{\ddag}$}~\affiliation{Brown University, Providence, Rhode Island 02912, USA}
\author{R.~Nayyar$^{\ddag}$}~\affiliation{University of Arizona, Tucson, Arizona 85721, USA}
\author{H.A.~Neal$^{\ddag}$}~\affiliation{University of Michigan, Ann Arbor, Michigan 48109, USA}
\author{J.P.~Negret$^{\ddag}$}~\affiliation{Universidad de los Andes, Bogot\'a, Colombia}
\author{J.~Nett$^{\dag}$}~\affiliation{Texas A\&M University, College Station, Texas 77843, USA}
\author{M.S.~Neubauer$^{\dag}$}~\affiliation{University of Illinois, Urbana, Illinois 61801, USA}
\author{C.~Neu$^{\dag}$}~\affiliation{University of Virginia, Charlottesville, Virginia 22904, USA}
\author{P.~Neustroev$^{\ddag}$}~\affiliation{Petersburg Nuclear Physics Institute, St. Petersburg, Russia}
\author{H.T.~Nguyen$^{\ddag}$} \affiliation{University of Virginia, Charlottesville, Virginia 22904, USA}
\author{J.~Nielsen$^{\dag v}$}~\affiliation{Ernest Orlando Lawrence Berkeley National Laboratory, Berkeley, California 94720, USA}
\author{L.~Nodulman$^{\dag}$}~\affiliation{Argonne National Laboratory, Argonne, Illinois 60439, USA}
\author{S.Y.~Noh$^{\dag}$}~\affiliation{Center for High Energy Physics: Kyungpook National University, Daegu 702-701, Korea; Seoul National University, Seoul 151-742, Korea; Sungkyunkwan University, Suwon 440-746, Korea; Korea Institute of Science and Technology Information, Daejeon 305-806, Korea; Chonnam National University, Gwangju 500-757, Korea; Chonbuk National University, Jeonju 561-756, Korea}
\author{O.~Norniella$^{\dag}$}~\affiliation{University of Illinois, Urbana, Illinois 61801, USA}
\author{T.~Nunnemann$^{\ddag}$}~\affiliation{Ludwig-Maximilians-Universit\"at M\"unchen, M\"unchen, Germany}
\author{L.~Oakes$^{\dag}$}~\affiliation{University of Oxford, Oxford OX1 3RH, United Kingdom}
\author{S.H.~Oh$^{\dag}$}~\affiliation{Duke University, Durham, North Carolina 27708, USA}
\author{Y.D.~Oh$^{\dag}$}~\affiliation{Center for High Energy Physics: Kyungpook National University, Daegu 702-701, Korea; Seoul National University, Seoul 151-742, Korea; Sungkyunkwan University, Suwon 440-746, Korea; Korea Institute of Science and Technology Information, Daejeon 305-806, Korea; Chonnam National University, Gwangju 500-757, Korea; Chonbuk National University, Jeonju 561-756, Korea}
\author{I.~Oksuzian$^{\dag}$}~\affiliation{University of Virginia, Charlottesville, Virginia 22904, USA}
\author{T.~Okusawa$^{\dag}$}~\affiliation{Osaka City University, Osaka 588, Japan}
\author{R.~Orava$^{\dag}$}~\affiliation{Division of High Energy Physics, Department of Physics, University of Helsinki and Helsinki Institute of Physics, FIN-00014, Helsinki, Finland}
\author{J.~Orduna$^{\ddag}$}~\affiliation{Rice University, Houston, Texas 77005, USA}
\author{L.~Ortolan$^{\dag}$}~\affiliation{Institut de Fisica d'Altes Energies, ICREA, Universitat Autonoma de Barcelona, E-08193, Bellaterra (Barcelona), Spain}
\author{N.~Osman$^{\ddag}$}~\affiliation{CPPM, Aix-Marseille Universit\'e, CNRS/IN2P3, Marseille, France}
\author{J.~Osta$^{\ddag}$}~\affiliation{University of Notre Dame, Notre Dame, Indiana 46556, USA}
\author{M.~Padilla$^{\ddag}$}~\affiliation{University of California Riverside, Riverside, California 92521, USA}
\author{S.~Pagan~Griso$^{\dag\sharp{b}}$}~\affiliation{Istituto Nazionale di Fisica Nucleare, Sezione di Padova-Trento, $^{\sharp{b}}$University of Padova, I-35131 Padova, Italy}
\author{C.~Pagliarone$^{\dag}$}~\affiliation{Istituto Nazionale di Fisica Nucleare Trieste/Udine, I-34100 Trieste, $^{\sharp{g}}$University of Udine, I-33100 Udine, Italy}
\author{A.~Pal$^{\ddag}$}~\affiliation{University of Texas, Arlington, Texas 76019, USA}
\author{E.~Palencia$^{\dag e}$}~\affiliation{Instituto de Fisica de Cantabria, CSIC-University of Cantabria, 39005 Santander, Spain}
\author{V.~Papadimitriou$^{\dag}$}~\affiliation{Fermi National Accelerator Laboratory, Batavia, Illinois 60510, USA}
\author{A.A.~Paramonov$^{\dag}$}~\affiliation{Argonne National Laboratory, Argonne, Illinois 60439, USA}
\author{N.~Parashar$^{\ddag}$}~\affiliation{Purdue University Calumet, Hammond, Indiana 46323, USA}
\author{V.~Parihar$^{\ddag}$}~\affiliation{Brown University, Providence, Rhode Island 02912, USA}
\author{S.K.~Park$^{\ddag}$}~\affiliation{Korea Detector Laboratory, Korea University, Seoul, Korea}
\author{R.~Partridge$^{\ddag g}$}~\affiliation{Brown University, Providence, Rhode Island 02912, USA}
\author{N.~Parua$^{\ddag}$}~\affiliation{Indiana University, Bloomington, Indiana 47405, USA}
\author{J.~Patrick$^{\dag}$}~\affiliation{Fermi National Accelerator Laboratory, Batavia, Illinois 60510, USA}
\author{A.~Patwa$^{\ddag}$}~\affiliation{Brookhaven National Laboratory, Upton, New York 11973, USA}
\author{G.~Pauletta$^{\dag\sharp{g}}$}~\affiliation{Istituto Nazionale di Fisica Nucleare Trieste/Udine, I-34100 Trieste, $^{\sharp{g}}$University of Udine, I-33100 Udine, Italy}
\author{M.~Paulini$^{\dag}$}~\affiliation{Carnegie Mellon University, Pittsburgh, Pennsylvania 15213, USA}
\author{C.~Paus$^{\dag}$}~\affiliation{Massachusetts Institute of Technology, Cambridge, Massachusetts 02139, USA}
\author{D.E.~Pellett$^{\dag}$}~\affiliation{University of California, Davis, Davis, California 95616, USA}
\author{B.~Penning$^{\ddag}$}~\affiliation{Fermi National Accelerator Laboratory, Batavia, Illinois 60510, USA}
\author{A.~Penzo$^{\dag}$}~\affiliation{Istituto Nazionale di Fisica Nucleare Trieste/Udine, I-34100 Trieste, $^{\sharp{g}}$University of Udine, I-33100 Udine, Italy}
\author{M.~Perfilov$^{\ddag}$}~\affiliation{Moscow State University, Moscow, Russia}
\author{Y.~Peters$^{\ddag}$}~\affiliation{The University of Manchester, Manchester M13 9PL, United Kingdom}
\author{K.~Petridis$^{\ddag}$}~\affiliation{The University of Manchester, Manchester M13 9PL, United Kingdom}
\author{G.~Petrillo$^{\ddag}$}~\affiliation{University of Rochester, Rochester, New York 14627, USA}
\author{P.~P\'etroff$^{\ddag}$}~\affiliation{LAL, Universit\'e Paris-Sud, CNRS/IN2P3, Orsay, France}
\author{T.J.~Phillips$^{\dag}$}~\affiliation{Duke University, Durham, North Carolina 27708, USA}
\author{G.~Piacentino$^{\dag}$}~\affiliation{Istituto Nazionale di Fisica Nucleare Pisa, $^{\sharp{c}}$University of Pisa, $^{\sharp{d}}$University of Siena and $^{\sharp{e}}$Scuola Normale Superiore, I-56127 Pisa, Italy}
\author{E.~Pianori$^{\dag}$}~\affiliation{University of Pennsylvania, Philadelphia, Pennsylvania 19104, USA}
\author{J.~Pilot$^{\dag}$}~\affiliation{The Ohio State University, Columbus, Ohio 43210, USA}
\author{K.~Pitts$^{\dag}$}~\affiliation{University of Illinois, Urbana, Illinois 61801, USA}
\author{C.~Plager$^{\dag}$}~\affiliation{University of California, Los Angeles, Los Angeles, California 90024, USA}
\author{M.-A.~Pleier$^{\ddag}$}~\affiliation{Brookhaven National Laboratory, Upton, New York 11973, USA}
\author{P.L.M.~Podesta-Lerma$^{\ddag h}$}~\affiliation{CINVESTAV, Mexico City, Mexico}
\author{V.M.~Podstavkov$^{\ddag}$}~\affiliation{Fermi National Accelerator Laboratory, Batavia, Illinois 60510, USA}
\author{L.~Pondrom$^{\dag}$}~\affiliation{University of Wisconsin, Madison, Wisconsin 53706, USA}
\author{A.V.~Popov$^{\ddag}$}~\affiliation{Institute for High Energy Physics, Protvino, Russia}
\author{S.~Poprocki$^{\dag j}$}~\affiliation{Fermi National Accelerator Laboratory, Batavia, Illinois 60510, USA}
\author{K.~Potamianos$^{\dag}$}~\affiliation{Purdue University, West Lafayette, Indiana 47907, USA}
\author{A.~Pranko$^{\dag}$}~\affiliation{Ernest Orlando Lawrence Berkeley National Laboratory, Berkeley, California 94720, USA}
\author{M.~Prewitt$^{\ddag}$}~\affiliation{Rice University, Houston, Texas 77005, USA}
\author{D.~Price$^{\ddag}$}~\affiliation{Indiana University, Bloomington, Indiana 47405, USA}
\author{N.~Prokopenko$^{\ddag}$}~\affiliation{Institute for High Energy Physics, Protvino, Russia}
\author{F.~Prokoshin$^{\dag w}$}~\affiliation{Joint Institute for Nuclear Research, Dubna, Russia}
\author{F.~Ptohos$^{\dag x}$}~\affiliation{Laboratori Nazionali di Frascati, Istituto Nazionale di Fisica Nucleare, I-00044 Frascati, Italy}
\author{G.~Punzi$^{\dag\sharp{c}}$}~\affiliation{Istituto Nazionale di Fisica Nucleare Pisa, $^{\sharp{c}}$University of Pisa, $^{\sharp{d}}$University of Siena and $^{\sharp{e}}$Scuola Normale Superiore, I-56127 Pisa, Italy}
\author{J.~Qian$^{\ddag}$}~\affiliation{University of Michigan, Ann Arbor, Michigan 48109, USA}
\author{A.~Quadt$^{\ddag}$}~\affiliation{II. Physikalisches Institut, Georg-August-Universit\"at G\"ottingen, G\"ottingen, Germany}
\author{B.~Quinn$^{\ddag}$}~\affiliation{University of Mississippi, University, Mississippi 38677, USA}
\author{A.~Rahaman$^{\dag}$}~\affiliation{University of Pittsburgh, Pittsburgh, Pennsylvania 15260, USA}
\author{V.~Ramakrishnan$^{\dag}$}~\affiliation{University of Wisconsin, Madison, Wisconsin 53706, USA}
\author{M.S.~Rangel$^{\ddag}$}~\affiliation{LAFEX, Centro Brasileiro de Pesquisas F\'{i}sicas, Rio de Janeiro, Brazil}
\author{K.~Ranjan$^{\ddag}$}~\affiliation{Delhi University, Delhi, India}
\author{N.~Ranjan$^{\dag}$}~\affiliation{Purdue University, West Lafayette, Indiana 47907, USA}
\author{P.N.~Ratoff$^{\ddag}$}~\affiliation{Lancaster University, Lancaster LA1 4YB, United Kingdom}
\author{I.~Razumov$^{\ddag}$}~\affiliation{Institute for High Energy Physics, Protvino, Russia}
\author{I.~Redondo$^{\dag}$}~\affiliation{Centro de Investigaciones Energeticas Medioambientales y Tecnologicas, E-28040 Madrid, Spain}
\author{P.~Renkel$^{\ddag}$}~\affiliation{Southern Methodist University, Dallas, Texas 75275, USA}
\author{P.~Renton$^{\dag}$}~\affiliation{University of Oxford, Oxford OX1 3RH, United Kingdom}
\author{M.~Rescigno$^{\dag}$}~\affiliation{Istituto Nazionale di Fisica Nucleare, Sezione di Roma 1, $^{\sharp{f}}$Sapienza Universit\`{a} di Roma, I-00185 Roma, Italy}
\author{T.~Riddick$^{\dag}$}~\affiliation{University College London, London WC1E 6BT, United Kingdom}
\author{F.~Rimondi$^{\dag\sharp{a}}$}~\affiliation{Istituto Nazionale di Fisica Nucleare Bologna, $^{\sharp{a}}$University of Bologna, I-40127 Bologna, Italy}
\author{I.~Ripp-Baudot$^{\ddag}$}~\affiliation{IPHC, Universit\'e de Strasbourg, CNRS/IN2P3, Strasbourg, France}
\author{L.~Ristori$^{\dag}$}\affiliation{Istituto Nazionale di Fisica Nucleare Pisa, $^{\sharp{c}}$University of Pisa, $^{\sharp{d}}$University of Siena and $^{\sharp{e}}$Scuola Normale Superiore, I-56127 Pisa, Italy}~\affiliation{Fermi National Accelerator Laboratory, Batavia, Illinois 60510, USA}
\author{F.~Rizatdinova$^{\ddag}$}~\affiliation{Oklahoma State University, Stillwater, Oklahoma 74078, USA}
\author{A.~Robson$^{\dag}$}~\affiliation{Glasgow University, Glasgow G12 8QQ, United Kingdom}
\author{T.~Rodriguez$^{\dag}$}~\affiliation{University of Pennsylvania, Philadelphia, Pennsylvania 19104, USA}
\author{E.~Rogers$^{\dag}$}~\affiliation{University of Illinois, Urbana, Illinois 61801, USA}
\author{S.~Rolli$^{\dag y}$}~\affiliation{Tufts University, Medford, Massachusetts 02155, USA}
\author{M.~Rominsky$^{\ddag}$}~\affiliation{Fermi National Accelerator Laboratory, Batavia, Illinois 60510, USA}
\author{R.~Roser$^{\dag}$}~\affiliation{Fermi National Accelerator Laboratory, Batavia, Illinois 60510, USA}
\author{A.~Ross$^{\ddag}$}~\affiliation{Lancaster University, Lancaster LA1 4YB, United Kingdom}
\author{C.~Royon$^{\ddag}$}~\affiliation{CEA, Irfu, SPP, Saclay, France}
\author{P.~Rubinov$^{\ddag}$}~\affiliation{Fermi National Accelerator Laboratory, Batavia, Illinois 60510, USA}
\author{R.~Ruchti$^{\ddag}$}~\affiliation{University of Notre Dame, Notre Dame, Indiana 46556, USA}
\author{F.~Ruffini$^{\dag\sharp{d}}$}~\affiliation{Istituto Nazionale di Fisica Nucleare Pisa, $^{\sharp{c}}$University of Pisa, $^{\sharp{d}}$University of Siena and $^{\sharp{e}}$Scuola Normale Superiore, I-56127 Pisa, Italy}
\author{A.~Ruiz$^{\dag}$}~\affiliation{Instituto de Fisica de Cantabria, CSIC-University of Cantabria, 39005 Santander, Spain}
\author{J.~Russ$^{\dag}$}~\affiliation{Carnegie Mellon University, Pittsburgh, Pennsylvania 15213, USA}
\author{V.~Rusu$^{\dag}$}~\affiliation{Fermi National Accelerator Laboratory, Batavia, Illinois 60510, USA}
\author{A.~Safonov$^{\dag}$}~\affiliation{Texas A\&M University, College Station, Texas 77843, USA}
\author{G.~Sajot$^{\ddag}$}~\affiliation{LPSC, Universit\'e Joseph Fourier Grenoble 1, CNRS/IN2P3, Institut National Polytechnique de Grenoble, Grenoble, France}
\author{W.K.~Sakumoto$^{\dag}$}~\affiliation{University of Rochester, Rochester, New York 14627, USA}
\author{Y.~Sakurai$^{\dag}$}~\affiliation{Waseda University, Tokyo 169, Japan}
\author{P.~Salcido$^{\ddag}$}~\affiliation{Northern Illinois University, DeKalb, Illinois 60115, USA}
\author{A.~S\'anchez-Hern\'andez$^{\ddag}$}~\affiliation{CINVESTAV, Mexico City, Mexico}
\author{M.P.~Sanders$^{\ddag}$}~\affiliation{Ludwig-Maximilians-Universit\"at M\"unchen, M\"unchen, Germany}
\author{L.~Santi$^{\dag\sharp{g}}$}~\affiliation{Istituto Nazionale di Fisica Nucleare Trieste/Udine, I-34100 Trieste, $^{\sharp{g}}$University of Udine, I-33100 Udine, Italy}
\author{A.S.~Santos$^{\ddag i}$}~\affiliation{LAFEX, Centro Brasileiro de Pesquisas F\'{i}sicas, Rio de Janeiro, Brazil}
\author{K.~Sato$^{\dag}$}~\affiliation{University of Tsukuba, Tsukuba, Ibaraki 305, Japan}
\author{G.~Savage$^{\ddag}$}~\affiliation{Fermi National Accelerator Laboratory, Batavia, Illinois 60510, USA}
\author{V.~Saveliev$^{\dag g}$}~\affiliation{Fermi National Accelerator Laboratory, Batavia, Illinois 60510, USA}
\author{A.~Savoy-Navarro$^{\dag z}$}~\affiliation{Fermi National Accelerator Laboratory, Batavia, Illinois 60510, USA}
\author{L.~Sawyer$^{\ddag}$}~\affiliation{Louisiana Tech University, Ruston, Louisiana 71272, USA}
\author{T.~Scanlon$^{\ddag}$}~\affiliation{Imperial College London, London SW7 2AZ, United Kingdom}
\author{R.D.~Schamberger$^{\ddag}$}~\affiliation{State University of New York, Stony Brook, New York 11794, USA}
\author{Y.~Scheglov$^{\ddag}$}~\affiliation{Petersburg Nuclear Physics Institute, St. Petersburg, Russia}
\author{H.~Schellman$^{\ddag}$}~\affiliation{Northwestern University, Evanston, Illinois 60208, USA}
\author{P.~Schlabach$^{\dag}$}~\affiliation{Fermi National Accelerator Laboratory, Batavia, Illinois 60510, USA}
\author{S.~Schlobohm$^{\ddag}$}~\affiliation{University of Washington, Seattle, Washington 98195, USA}
\author{A.~Schmidt$^{\dag}$}~\affiliation{Institut f\"{u}r Experimentelle Kernphysik, Karlsruhe Institute of Technology, D-76131 Karlsruhe, Germany}
\author{E.E.~Schmidt$^{\dag}$}~\affiliation{Fermi National Accelerator Laboratory, Batavia, Illinois 60510, USA}
\author{C.~Schwanenberger$^{\ddag}$}~\affiliation{The University of Manchester, Manchester M13 9PL, United Kingdom}
\author{T.~Schwarz$^{\dag}$}~\affiliation{Fermi National Accelerator Laboratory, Batavia, Illinois 60510, USA}
\author{R.~Schwienhorst$^{\ddag}$}~\affiliation{Michigan State University, East Lansing, Michigan 48824, USA}
\author{L.~Scodellaro$^{\dag}$}~\affiliation{Instituto de Fisica de Cantabria, CSIC-University of Cantabria, 39005 Santander, Spain}
\author{A.~Scribano$^{\dag\sharp{d}}$}~\affiliation{Istituto Nazionale di Fisica Nucleare Pisa, $^{\sharp{c}}$University of Pisa, $^{\sharp{d}}$University of Siena and $^{\sharp{e}}$Scuola Normale Superiore, I-56127 Pisa, Italy}
\author{F.~Scuri$^{\dag}$}~\affiliation{Istituto Nazionale di Fisica Nucleare Pisa, $^{\sharp{c}}$University of Pisa, $^{\sharp{d}}$University of Siena and $^{\sharp{e}}$Scuola Normale Superiore, I-56127 Pisa, Italy}
\author{S.~Seidel$^{\dag}$}~\affiliation{University of New Mexico, Albuquerque, New Mexico 87131, USA}
\author{Y.~Seiya$^{\dag}$}~\affiliation{Osaka City University, Osaka 588, Japan}
\author{J.~Sekaric$^{\ddag}$}~\affiliation{University of Kansas, Lawrence, Kansas 66045, USA}
\author{A.~Semenov$^{\dag}$}~\affiliation{Joint Institute for Nuclear Research, Dubna, Russia}
\author{H.~Severini$^{\ddag}$}~\affiliation{University of Oklahoma, Norman, Oklahoma 73019, USA}
\author{F.~Sforza$^{\dag\sharp{c}}$}~\affiliation{Istituto Nazionale di Fisica Nucleare Pisa, $^{\sharp{c}}$University of Pisa, $^{\sharp{d}}$University of Siena and $^{\sharp{e}}$Scuola Normale Superiore, I-56127 Pisa, Italy}
\author{E.~Shabalina$^{\ddag}$}~\affiliation{II. Physikalisches Institut, Georg-August-Universit\"at G\"ottingen, G\"ottingen, Germany}
\author{S.Z.~Shalhout$^{\dag}$}~\affiliation{University of California, Davis, Davis, California 95616, USA}
\author{V.~Shary$^{\ddag}$}~\affiliation{CEA, Irfu, SPP, Saclay, France}
\author{S.~Shaw$^{\ddag}$}~\affiliation{Michigan State University, East Lansing, Michigan 48824, USA}
\author{A.A.~Shchukin$^{\ddag}$}~\affiliation{Institute for High Energy Physics, Protvino, Russia}
\author{T.~Shears$^{\dag}$}~\affiliation{University of Liverpool, Liverpool L69 7ZE, United Kingdom}
\author{P.F.~Shepard$^{\dag}$}~\affiliation{University of Pittsburgh, Pittsburgh, Pennsylvania 15260, USA}
\author{M.~Shimojima$^{\dag aa}$}~\affiliation{University of Tsukuba, Tsukuba, Ibaraki 305, Japan}
\author{R.K.~Shivpuri$^{\ddag}$}~\affiliation{Delhi University, Delhi, India}
\author{M.~Shochet$^{\dag}$}~\affiliation{Enrico Fermi Institute, University of Chicago, Chicago, Illinois 60637, USA}
\author{I.~Shreyber-Tecker$^{\dag}$}~\affiliation{Institution for Theoretical and Experimental Physics, ITEP, Moscow 117259, Russia}
\author{V.~Simak$^{\ddag}$}~\affiliation{Czech Technical University in Prague, Prague, Czech Republic}
\author{A.~Simonenko$^{\dag}$}~\affiliation{Joint Institute for Nuclear Research, Dubna, Russia}
\author{P.~Sinervo$^{\dag}$}~\affiliation{Institute of Particle Physics: McGill University, Montr\'{e}al, Qu\'{e}bec, Canada H3A~2T8; Simon Fraser University, Burnaby, British Columbia, Canada V5A~1S6; University of Toronto, Toronto, Ontario, Canada M5S~1A7; and TRIUMF, Vancouver, British Columbia, Canada V6T~2A3}
\author{P.~Skubic$^{\ddag}$}~\affiliation{University of Oklahoma, Norman, Oklahoma 73019, USA}
\author{P.~Slattery$^{\ddag}$}~\affiliation{University of Rochester, Rochester, New York 14627, USA}
\author{K.~Sliwa$^{\dag}$}~\affiliation{Tufts University, Medford, Massachusetts 02155, USA}
\author{D.~Smirnov$^{\ddag}$}~\affiliation{University of Notre Dame, Notre Dame, Indiana 46556, USA}
\author{J.R.~Smith$^{\dag}$}~\affiliation{University of California, Davis, Davis, California 95616, USA}
\author{K.J.~Smith$^{\ddag}$}~\affiliation{State University of New York, Buffalo, New York 14260, USA}
\author{F.D.~Snider$^{\dag}$}~\affiliation{Fermi National Accelerator Laboratory, Batavia, Illinois 60510, USA}
\author{G.R.~Snow$^{\ddag}$}~\affiliation{University of Nebraska, Lincoln, Nebraska 68588, USA}
\author{J.~Snow$^{\ddag}$}~\affiliation{Langston University, Langston, Oklahoma 73050, USA}
\author{S.~Snyder$^{\ddag}$}~\affiliation{Brookhaven National Laboratory, Upton, New York 11973, USA}
\author{A.~Soha$^{\dag}$}~\affiliation{Fermi National Accelerator Laboratory, Batavia, Illinois 60510, USA}
\author{S.~S{\"o}ldner-Rembold$^{\ddag}$}~\affiliation{The University of Manchester, Manchester M13 9PL, United Kingdom}
\author{H.~Song$^{\dag}$}~\affiliation{University of Pittsburgh, Pittsburgh, Pennsylvania 15260, USA}
\author{L.~Sonnenschein$^{\ddag}$}~\affiliation{III. Physikalisches Institut A, RWTH Aachen University, Aachen, Germany}
\author{V.~Sorin$^{\dag}$}~\affiliation{Institut de Fisica d'Altes Energies, ICREA, Universitat Autonoma de Barcelona, E-08193, Bellaterra (Barcelona), Spain}
\author{K.~Soustruznik$^{\ddag}$}~\affiliation{Charles University, Faculty of Mathematics and Physics, Center for Particle Physics, Prague, Czech Republic}
\author{P.~Squillacioti$^{\dag\sharp{d}}$}~\affiliation{Istituto Nazionale di Fisica Nucleare Pisa, $^{\sharp{c}}$University of Pisa, $^{\sharp{d}}$University of Siena and $^{\sharp{e}}$Scuola Normale Superiore, I-56127 Pisa, Italy}
\author{R.~St.~Denis$^{\dag}$}~\affiliation{Glasgow University, Glasgow G12 8QQ, United Kingdom}
\author{M.~Stancari$^{\dag}$}~\affiliation{Fermi National Accelerator Laboratory, Batavia, Illinois 60510, USA}
\author{J.~Stark$^{\ddag}$}~\affiliation{LPSC, Universit\'e Joseph Fourier Grenoble 1, CNRS/IN2P3, Institut National Polytechnique de Grenoble, Grenoble, France}
\author{O.~Stelzer-Chilton$^{\dag}$}~\affiliation{Institute of Particle Physics: McGill University, Montr\'{e}al, Qu\'{e}bec, Canada H3A~2T8; Simon Fraser University, Burnaby, British Columbia, Canada V5A~1S6; University of Toronto, Toronto, Ontario, Canada M5S~1A7; and TRIUMF, Vancouver, British Columbia, Canada V6T~2A3}
\author{B.~Stelzer$^{\dag}$}~\affiliation{Institute of Particle Physics: McGill University, Montr\'{e}al, Qu\'{e}bec, Canada H3A~2T8; Simon Fraser University, Burnaby, British Columbia, Canada V5A~1S6; University of Toronto, Toronto, Ontario, Canada M5S~1A7; and TRIUMF, Vancouver, British Columbia, Canada V6T~2A3}
\author{D.~Stentz$^{\dag b}$}~\affiliation{Fermi National Accelerator Laboratory, Batavia, Illinois 60510, USA}
\author{D.A.~Stoyanova$^{\ddag}$}~\affiliation{Institute for High Energy Physics, Protvino, Russia}
\author{M.~Strauss$^{\ddag}$}~\affiliation{University of Oklahoma, Norman, Oklahoma 73019, USA}
\author{J.~Strologas$^{\dag}$}~\affiliation{University of New Mexico, Albuquerque, New Mexico 87131, USA}
\author{G.L.~Strycker$^{\dag}$}~\affiliation{University of Michigan, Ann Arbor, Michigan 48109, USA}
\author{Y.~Sudo$^{\dag}$}~\affiliation{University of Tsukuba, Tsukuba, Ibaraki 305, Japan}
\author{A.~Sukhanov$^{\dag}$}~\affiliation{Fermi National Accelerator Laboratory, Batavia, Illinois 60510, USA}
\author{I.~Suslov$^{\dag}$}~\affiliation{Joint Institute for Nuclear Research, Dubna, Russia}
\author{L.~Suter$^{\ddag}$}~\affiliation{The University of Manchester, Manchester M13 9PL, United Kingdom}
\author{P.~Svoisky$^{\ddag}$}~\affiliation{University of Oklahoma, Norman, Oklahoma 73019, USA}
\author{M.~Takahashi$^{\ddag}$}~\affiliation{The University of Manchester, Manchester M13 9PL, United Kingdom}
\author{K.~Takemasa$^{\dag}$}~\affiliation{University of Tsukuba, Tsukuba, Ibaraki 305, Japan}
\author{Y.~Takeuchi$^{\dag}$}~\affiliation{University of Tsukuba, Tsukuba, Ibaraki 305, Japan}
\author{J.~Tang$^{\dag}$}~\affiliation{Enrico Fermi Institute, University of Chicago, Chicago, Illinois 60637, USA}
\author{M.~Tecchio$^{\dag}$}~\affiliation{University of Michigan, Ann Arbor, Michigan 48109, USA}
\author{P.K.~Teng$^{\dag}$}~\affiliation{Institute of Physics, Academia Sinica, Taipei, Taiwan 11529, Republic of China}
\author{J.~Thom$^{\dag j}$}~\affiliation{Fermi National Accelerator Laboratory, Batavia, Illinois 60510, USA}
\author{J.~Thome$^{\dag}$}~\affiliation{Carnegie Mellon University, Pittsburgh, Pennsylvania 15213, USA}
\author{G.A.~Thompson$^{\dag}$}~\affiliation{University of Illinois, Urbana, Illinois 61801, USA}
\author{E.~Thomson$^{\dag}$}~\affiliation{University of Pennsylvania, Philadelphia, Pennsylvania 19104, USA}
\author{M.~Titov$^{\ddag}$}~\affiliation{CEA, Irfu, SPP, Saclay, France}
\author{D.~Toback$^{\dag}$}~\affiliation{Texas A\&M University, College Station, Texas 77843, USA}
\author{S.~Tokar$^{\dag}$}~\affiliation{Comenius University, 842 48 Bratislava, Slovakia; Institute of Experimental Physics, 040 01 Kosice, Slovakia}
\author{V.V.~Tokmenin$^{\ddag}$}~\affiliation{Joint Institute for Nuclear Research, Dubna, Russia}
\author{K.~Tollefson$^{\dag}$}~\affiliation{Michigan State University, East Lansing, Michigan 48824, USA}
\author{T.~Tomura$^{\dag}$}~\affiliation{University of Tsukuba, Tsukuba, Ibaraki 305, Japan}
\author{D.~Tonelli$^{\dag}$}~\affiliation{Fermi National Accelerator Laboratory, Batavia, Illinois 60510, USA}
\author{S.~Torre$^{\dag}$}~\affiliation{Laboratori Nazionali di Frascati, Istituto Nazionale di Fisica Nucleare, I-00044 Frascati, Italy}
\author{D.~Torretta$^{\dag}$}~\affiliation{Fermi National Accelerator Laboratory, Batavia, Illinois 60510, USA}
\author{P.~Totaro$^{\dag}$}~\affiliation{Istituto Nazionale di Fisica Nucleare, Sezione di Padova-Trento, $^{\sharp{b}}$University of Padova, I-35131 Padova, Italy}
\author{M.~Trovato$^{\dag\sharp{e}}$}~\affiliation{Istituto Nazionale di Fisica Nucleare Pisa, $^{\sharp{c}}$University of Pisa, $^{\sharp{d}}$University of Siena and $^{\sharp{e}}$Scuola Normale Superiore, I-56127 Pisa, Italy}
\author{Y.-T.~Tsai$^{\ddag}$}~\affiliation{University of Rochester, Rochester, New York 14627, USA}
\author{K.~Tschann-Grimm$^{\ddag}$}~\affiliation{State University of New York, Stony Brook, New York 11794, USA}
\author{D.~Tsybychev$^{\ddag}$}~\affiliation{State University of New York, Stony Brook, New York 11794, USA}
\author{B.~Tuchming$^{\ddag}$}~\affiliation{CEA, Irfu, SPP, Saclay, France}
\author{C.~Tully$^{\ddag}$}~\affiliation{Princeton University, Princeton, New Jersey 08544, USA}
\author{F.~Ukegawa$^{\dag}$}~\affiliation{University of Tsukuba, Tsukuba, Ibaraki 305, Japan}
\author{S.~Uozumi$^{\dag}$}~\affiliation{Center for High Energy Physics: Kyungpook National University, Daegu 702-701, Korea; Seoul National University, Seoul 151-742, Korea; Sungkyunkwan University, Suwon 440-746, Korea; Korea Institute of Science and Technology Information, Daejeon 305-806, Korea; Chonnam National University, Gwangju 500-757, Korea; Chonbuk National University, Jeonju 561-756, Korea}
\author{L.~Uvarov$^{\ddag}$}~\affiliation{Petersburg Nuclear Physics Institute, St. Petersburg, Russia}
\author{S.~Uvarov$^{\ddag}$}~\affiliation{Petersburg Nuclear Physics Institute, St. Petersburg, Russia}
\author{S.~Uzunyan$^{\ddag}$}~\affiliation{Northern Illinois University, DeKalb, Illinois 60115, USA}
\author{R.~Van~Kooten$^{\ddag}$}~\affiliation{Indiana University, Bloomington, Indiana 47405, USA}
\author{W.M.~van~Leeuwen$^{\ddag}$}~\affiliation{Nikhef, Science Park, Amsterdam, the Netherlands}
\author{N.~Varelas$^{\ddag}$}~\affiliation{University of Illinois at Chicago, Chicago, Illinois 60607, USA}
\author{A.~Varganov$^{\dag}$}~\affiliation{University of Michigan, Ann Arbor, Michigan 48109, USA}
\author{E.W.~Varnes$^{\ddag}$}~\affiliation{University of Arizona, Tucson, Arizona 85721, USA}
\author{I.A.~Vasilyev$^{\ddag}$}~\affiliation{Institute for High Energy Physics, Protvino, Russia}
\author{F.~V\'{a}zquez$^{\dag c}$}~\affiliation{University of Florida, Gainesville, Florida 32611, USA}
\author{G.~Velev$^{\dag}$}~\affiliation{Fermi National Accelerator Laboratory, Batavia, Illinois 60510, USA}
\author{C.~Vellidis$^{\dag}$}~\affiliation{Fermi National Accelerator Laboratory, Batavia, Illinois 60510, USA}
\author{P.~Verdier$^{\ddag}$}~\affiliation{IPNL, Universit\'e Lyon 1, CNRS/IN2P3, Villeurbanne, France and Universit\'e de Lyon, Lyon, France}
\author{A.Y.~Verkheev$^{\ddag}$}~\affiliation{Joint Institute for Nuclear Research, Dubna, Russia}
\author{L.S.~Vertogradov$^{\ddag}$}~\affiliation{Joint Institute for Nuclear Research, Dubna, Russia}
\author{M.~Verzocchi$^{\ddag}$}~\affiliation{Fermi National Accelerator Laboratory, Batavia, Illinois 60510, USA}
\author{M.~Vesterinen$^{\ddag}$}~\affiliation{The University of Manchester, Manchester M13 9PL, United Kingdom}
\author{M.~Vidal$^{\dag}$}~\affiliation{Purdue University, West Lafayette, Indiana 47907, USA}
\author{I.~Vila$^{\dag}$}~\affiliation{Instituto de Fisica de Cantabria, CSIC-University of Cantabria, 39005 Santander, Spain}
\author{D.~Vilanova$^{\ddag}$}~\affiliation{CEA, Irfu, SPP, Saclay, France}
\author{R.~Vilar$^{\dag}$}~\affiliation{Instituto de Fisica de Cantabria, CSIC-University of Cantabria, 39005 Santander, Spain}
\author{J.~Viz\'{a}n$^{\dag}$}~\affiliation{Instituto de Fisica de Cantabria, CSIC-University of Cantabria, 39005 Santander, Spain}
\author{M.~Vogel$^{\dag}$}~\affiliation{University of New Mexico, Albuquerque, New Mexico 87131, USA}
\author{P.~Vokac$^{\ddag}$}~\affiliation{Czech Technical University in Prague, Prague, Czech Republic}
\author{G.~Volpi$^{\dag}$}~\affiliation{Laboratori Nazionali di Frascati, Istituto Nazionale di Fisica Nucleare, I-00044 Frascati, Italy}
\author{P.~Wagner$^{\dag}$}~\affiliation{University of Pennsylvania, Philadelphia, Pennsylvania 19104, USA}
\author{R.L.~Wagner$^{\dag}$}~\affiliation{Fermi National Accelerator Laboratory, Batavia, Illinois 60510, USA}
\author{H.D.~Wahl$^{\ddag}$}~\affiliation{Florida State University, Tallahassee, Florida 32306, USA}
\author{T.~Wakisaka$^{\dag}$}~\affiliation{Osaka City University, Osaka 588, Japan}
\author{R.~Wallny$^{\dag}$}~\affiliation{University of California, Los Angeles, Los Angeles, California 90024, USA}
\author{S.M.~Wang$^{\dag}$}~\affiliation{Institute of Physics, Academia Sinica, Taipei, Taiwan 11529, Republic of China}
\author{M.H.L.S.~Wang$^{\ddag}$}~\affiliation{Fermi National Accelerator Laboratory, Batavia, Illinois 60510, USA}
\author{R.-J.~Wang$^{\ddag}$} \affiliation{Northeastern University, Boston, Massachusetts 02115, USA}
\author{A.~Warburton$^{\dag}$}~\affiliation{Institute of Particle Physics: McGill University, Montr\'{e}al, Qu\'{e}bec, Canada H3A~2T8; Simon Fraser University, Burnaby, British Columbia, Canada V5A~1S6; University of Toronto, Toronto, Ontario, Canada M5S~1A7; and TRIUMF, Vancouver, British Columbia, Canada V6T~2A3}
\author{J.~Warchol$^{\ddag}$}~\affiliation{University of Notre Dame, Notre Dame, Indiana 46556, USA}
\author{D.~Waters$^{\dag}$}~\affiliation{University College London, London WC1E 6BT, United Kingdom}
\author{G.~Watts$^{\ddag}$}~\affiliation{University of Washington, Seattle, Washington 98195, USA}
\author{M.~Wayne$^{\ddag}$}~\affiliation{University of Notre Dame, Notre Dame, Indiana 46556, USA}
\author{J.~Weichert$^{\ddag}$}~\affiliation{Institut f\"ur Physik, Universit\"at Mainz, Mainz, Germany}
\author{L.~Welty-Rieger$^{\ddag}$}~\affiliation{Northwestern University, Evanston, Illinois 60208, USA}
\author{W.C.~Wester~III$^{\dag}$}~\affiliation{Fermi National Accelerator Laboratory, Batavia, Illinois 60510, USA}
\author{A.~White$^{\ddag}$}~\affiliation{University of Texas, Arlington, Texas 76019, USA}
\author{D.~Whiteson$^{\dag bb}$}~\affiliation{University of Pennsylvania, Philadelphia, Pennsylvania 19104, USA}
\author{F.~Wick$^{\dag}$}~\affiliation{Institut f\"{u}r Experimentelle Kernphysik, Karlsruhe Institute of Technology, D-76131 Karlsruhe, Germany}
\author{D.~Wicke$^{\ddag}$}~\affiliation{Fachbereich Physik, Bergische Universit\"at Wuppertal, Wuppertal, Germany}
\author{A.B.~Wicklund$^{\dag}$}~\affiliation{Argonne National Laboratory, Argonne, Illinois 60439, USA}
\author{E.~Wicklund$^{\dag}$}~\affiliation{Fermi National Accelerator Laboratory, Batavia, Illinois 60510, USA}
\author{S.~Wilbur$^{\dag}$}~\affiliation{Enrico Fermi Institute, University of Chicago, Chicago, Illinois 60637, USA}
\author{H.H.~Williams$^{\dag}$}~\affiliation{University of Pennsylvania, Philadelphia, Pennsylvania 19104, USA}
\author{M.R.J.~Williams$^{\ddag}$}~\affiliation{Lancaster University, Lancaster LA1 4YB, United Kingdom}
\author{G.W.~Wilson$^{\ddag}$}~\affiliation{University of Kansas, Lawrence, Kansas 66045, USA}
\author{J.S.~Wilson$^{\dag}$}~\affiliation{The Ohio State University, Columbus, Ohio 43210, USA}
\author{P.~Wilson$^{\dag}$}~\affiliation{Fermi National Accelerator Laboratory, Batavia, Illinois 60510, USA}
\author{B.L.~Winer$^{\dag}$}~\affiliation{The Ohio State University, Columbus, Ohio 43210, USA}
\author{P.~Wittich$^{\dag j}$}~\affiliation{Fermi National Accelerator Laboratory, Batavia, Illinois 60510, USA}
\author{M.~Wobisch$^{\ddag}$}~\affiliation{Louisiana Tech University, Ruston, Louisiana 71272, USA}
\author{S.~Wolbers$^{\dag}$}~\affiliation{Fermi National Accelerator Laboratory, Batavia, Illinois 60510, USA}
\author{H.~Wolfe$^{\dag}$}~\affiliation{The Ohio State University, Columbus, Ohio 43210, USA}
\author{D.R.~Wood$^{\ddag}$}~\affiliation{Northeastern University, Boston, Massachusetts 02115, USA}
\author{T.~Wright$^{\dag}$}~\affiliation{University of Michigan, Ann Arbor, Michigan 48109, USA}
\author{X.~Wu$^{\dag}$}~\affiliation{University of Geneva, CH-1211 Geneva 4, Switzerland}
\author{Z.~Wu$^{\dag}$}~\affiliation{Baylor University, Waco, Texas 76798, USA}
\author{T.R.~Wyatt$^{\ddag}$}~\affiliation{The University of Manchester, Manchester M13 9PL, United Kingdom}
\author{Y.~Xie$^{\ddag}$}~\affiliation{Fermi National Accelerator Laboratory, Batavia, Illinois 60510, USA}
\author{R.~Yamada$^{\ddag}$}~\affiliation{Fermi National Accelerator Laboratory, Batavia, Illinois 60510, USA}
\author{K.~Yamamoto$^{\dag}$}~\affiliation{Osaka City University, Osaka 588, Japan}
\author{D.~Yamato$^{\dag}$}~\affiliation{Osaka City University, Osaka 588, Japan}
\author{S.~Yang$^{\ddag}$}~\affiliation{University of Science and Technology of China, Hefei, People's Republic of China}
\author{T.~Yang$^{\dag}$}~\affiliation{Fermi National Accelerator Laboratory, Batavia, Illinois 60510, USA}
\author{U.K.~Yang$^{\dag cc}$}~\affiliation{Enrico Fermi Institute, University of Chicago, Chicago, Illinois 60637, USA}
\author{W.-C.~Yang$^{\ddag}$}~\affiliation{The University of Manchester, Manchester M13 9PL, United Kingdom}
\author{Y.C.~Yang$^{\dag}$}~\affiliation{Center for High Energy Physics: Kyungpook National University, Daegu 702-701, Korea; Seoul National University, Seoul 151-742, Korea; Sungkyunkwan University, Suwon 440-746, Korea; Korea Institute of Science and Technology Information, Daejeon 305-806, Korea; Chonnam National University, Gwangju 500-757, Korea; Chonbuk National University, Jeonju 561-756, Korea}
\author{W.-M.~Yao$^{\dag}$}~\affiliation{Ernest Orlando Lawrence Berkeley National Laboratory, Berkeley, California 94720, USA}
\author{T.~Yasuda$^{\ddag}$}~\affiliation{Fermi National Accelerator Laboratory, Batavia, Illinois 60510, USA}
\author{Y.A.~Yatsunenko$^{\ddag}$}~\affiliation{Joint Institute for Nuclear Research, Dubna, Russia}
\author{W.~Ye$^{\ddag}$}~\affiliation{State University of New York, Stony Brook, New York 11794, USA}
\author{Z.~Ye$^{\ddag}$}~\affiliation{Fermi National Accelerator Laboratory, Batavia, Illinois 60510, USA}
\author{G.P.~Yeh$^{\dag}$}~\affiliation{Fermi National Accelerator Laboratory, Batavia, Illinois 60510, USA}
\author{K.~Yi$^{\dag u}$}~\affiliation{Fermi National Accelerator Laboratory, Batavia, Illinois 60510, USA}
\author{H.~Yin$^{\ddag}$}~\affiliation{Fermi National Accelerator Laboratory, Batavia, Illinois 60510, USA}
\author{K.~Yip$^{\ddag}$}~\affiliation{Brookhaven National Laboratory, Upton, New York 11973, USA}
\author{J.~Yoh$^{\dag}$}~\affiliation{Fermi National Accelerator Laboratory, Batavia, Illinois 60510, USA}
\author{K.~Yorita$^{\dag}$}~\affiliation{Waseda University, Tokyo 169, Japan}
\author{T.~Yoshida$^{\dag dd}$}~\affiliation{Osaka City University, Osaka 588, Japan}
\author{S.W.~Youn$^{\ddag}$}~\affiliation{Fermi National Accelerator Laboratory, Batavia, Illinois 60510, USA}
\author{G.B.~Yu$^{\dag}$}~\affiliation{Duke University, Durham, North Carolina 27708, USA}
\author{I.~Yu$^{\dag}$}~\affiliation{Center for High Energy Physics: Kyungpook National University, Daegu 702-701, Korea; Seoul National University, Seoul 151-742, Korea; Sungkyunkwan University, Suwon 440-746, Korea; Korea Institute of Science and Technology Information, Daejeon 305-806, Korea; Chonnam National University, Gwangju 500-757, Korea; Chonbuk National University, Jeonju 561-756, Korea}
\author{J.M.~Yu$^{\ddag}$}~\affiliation{University of Michigan, Ann Arbor, Michigan 48109, USA}
\author{S.S.~Yu$^{\dag}$}~\affiliation{Fermi National Accelerator Laboratory, Batavia, Illinois 60510, USA}
\author{J.C.~Yun$^{\dag}$}~\affiliation{Fermi National Accelerator Laboratory, Batavia, Illinois 60510, USA}
\author{A.~Zanetti$^{\dag}$}~\affiliation{Istituto Nazionale di Fisica Nucleare Trieste/Udine, I-34100 Trieste, $^{\sharp{g}}$University of Udine, I-33100 Udine, Italy}
\author{Y.~Zeng$^{\dag}$}~\affiliation{Duke University, Durham, North Carolina 27708, USA}
\author{J.~Zennamo$^{\ddag}$}~\affiliation{State University of New York, Buffalo, New York 14260, USA}
\author{T.~Zhao$^{\ddag}$}~\affiliation{University of Washington, Seattle, Washington 98195, USA}
\author{T.G.~Zhao$^{\ddag}$}~\affiliation{The University of Manchester, Manchester M13 9PL, United Kingdom}
\author{B.~Zhou$^{\ddag}$}~\affiliation{University of Michigan, Ann Arbor, Michigan 48109, USA}
\author{C.~Zhou$^{\dag}$}~\affiliation{Duke University, Durham, North Carolina 27708, USA}
\author{J.~Zhu$^{\ddag}$}~\affiliation{University of Michigan, Ann Arbor, Michigan 48109, USA}
\author{M.~Zielinski$^{\ddag}$}~\affiliation{University of Rochester, Rochester, New York 14627, USA}
\author{D.~Zieminska$^{\ddag}$}~\affiliation{Indiana University, Bloomington, Indiana 47405, USA}
\author{L.~Zivkovic$^{\ddag}$}~\affiliation{Brown University, Providence, Rhode Island 02912, USA}
\author{S.~Zucchelli$^{\dag\sharp{a}}$}~\affiliation{Istituto Nazionale di Fisica Nucleare Bologna, $^{\sharp{a}}$University of Bologna, I-40127 Bologna, Italy}
\collaboration{CDF\footnote{
With CDF visitors from
$^{{\dag}a}$Universidad de Oviedo, E-33007 Oviedo, Spain,
$^{{\dag}b}$Northwestern University, Evanston, IL 60208, USA,
$^{{\dag}c}$Universidad Iberoamericana, Mexico D.F., Mexico,
$^{{\dag}d}$ETH, 8092 Zurich, Switzerland,
$^{{\dag}e}$CERN, CH-1211 Geneva, Switzerland,
$^{{\dag}f}$Queen Mary, University of London, London, E1 4NS, United Kingdom,
$^{{\dag}g}$National Research Nuclear University, Moscow, Russia,
$^{{\dag}h}$Yarmouk University, Irbid 211-63, Jordan,
$^{{\dag}i}$Muons, Inc., Batavia, IL 60510, USA,
$^{{\dag}j}$Cornell University, Ithaca, NY 14853, USA,
$^{{\dag}k}$Kansas State University, Manhattan, KS 66506, USA,
$^{{\dag}l}$Kinki University, Higashi-Osaka City, Japan 577-8502,
$^{{\dag}m}$University of California, Santa Barbara, Santa Barbara, CA 93106, USA,
$^{{\dag}n}$University of Notre Dame, Notre Dame, IN 46556, USA,
$^{{\dag}o}$Ewha Womans University, Seoul, 120-750, Korea,
$^{{\dag}p}$Texas Tech University, Lubbock, TX 79609, USA,
$^{{\dag}q}$University of Melbourne, Victoria 3010, Australia,
$^{{\dag}r}$Institute of Physics, Academy of Sciences of the Czech Republic, Czech Republic,
$^{{\dag}s}$Istituto Nazionale di Fisica Nucleare, Sezione di Cagliari, 09042 Monserrato (Cagliari), Italy,
$^{{\dag}t}$University College Dublin, Dublin 4, Ireland,
$^{{\dag}u}$University of Iowa, Iowa City, IA 52242, USA,
$^{{\dag}v}$University of California, Santa Cruz, Santa Cruz, CA 95064, USA,
$^{{\dag}w}$Universidad Tecnica Federico Santa Maria, 110v Valparaiso, Chile,
$^{{\dag}x}$University of Cyprus, Nicosia CY-1678, Cyprus,
$^{{\dag}y}$Office of Science, U.S. Department of Energy, Washington, DC 20585, USA,
$^{{\dag}z}$CNRS-IN2P3, Paris, F-75205 France,
$^{{\dag}aa}$Nagasaki Institute of Applied Science, Nagasaki, Japan,
$^{{\dag}bb}$University of California, Irvine, Irvine, CA 92697, USA,
$^{{\dag}cc}$University of Manchester, Manchester M13 9PL, United Kingdom,
and 
$^{{\dag}dd}$University of Fukui, Fukui City, Fukui Prefecture, Japan 910-0017.
} and D0\footnote{
and D0 visitors from
%{alton}
$^{{\ddag}a}$Augustana College, Sioux Falls, SD, USA,
%{burdin}
$^{{\ddag}b}$The University of Liverpool, Liverpool, UK,
%{garcia-guerra}
$^{{\ddag}c}$UPIITA-IPN, Mexico City, Mexico,
%{grohsjean}
$^{{\ddag}d}$DESY, Hamburg, Germany,
%{hesketh}
$^{{\ddag}e}$University College London, London, UK,
%{luna-garcia}
$^{{\ddag}f}$Centro de Investigacion en Computacion - IPN, Mexico City, Mexico,
%{partridge}
$^{{\ddag}g}$SLAC, Menlo Park, CA, USA,
%{podesta-lerma}
$^{{\ddag}h}$ECFM, Universidad Autonoma de Sinaloa, Culiac\'an, Mexico
and
%{santos}
$^{{\ddag}i}$Universidade Estadual Paulista, S\~ao Paulo, Brazil.
} Collaborations}
\noaffiliation

\date{July 26, 2012}

\begin{abstract}

We combine searches by the CDF and D0 Collaborations for the associated production 
of a Higgs boson with a $W$ or $Z$ boson and subsequent decay of the Higgs boson to
a bottom-antibottom quark pair.  The data, originating from Fermilab Tevatron $p{\bar{p}}$ collisions 
at $\sqrt{s}=1.96$~TeV, correspond to integrated luminosities of up to 9.7 fb$^{-1}$.  The
searches are conducted for a Higgs boson with mass in the range 100--150~GeV$/c^2$.
We observe an excess of events in the data compared with the background predictions,
which is most significant in the mass range between 120 and 135~GeV/$c^2$.  The 
largest local significance is 3.3 standard deviations, corresponding to a global  
significance of 3.1 standard deviations.  We interpret this as evidence for the presence of
a new particle consistent with the standard model Higgs boson, which is produced in association with 
a weak vector boson and decays to a bottom-antibottom quark pair.

\end{abstract}

% activate the following line for publication

\maketitle

The standard model (SM)~\cite{gws,higgs} Higgs boson $H$ is predicted to be
produced in association with a $W$ or $Z$ boson at the Fermilab Tevatron 
$p{\bar{p}}$ Collider if it is within kinematic reach, and its dominant 
decay mode is predicted to be into a bottom-antibottom quark pair ($b{\bar{b}}$), 
if its mass $m_H$ is less than 135~GeV/$c^2$~\cite{vhtheory,lhcdifferential}.
An observation of this process would support
the SM prediction that the mechanism for electroweak symmetry breaking, which
gives mass to the weak vector bosons, is also the source of fermionic mass in the
quark sector.
The leptonic decays of the $W$ and $Z$ vector bosons and the 
decays of the $H$ to $b{\bar{b}}$ provide distinctive signatures of Higgs boson
production, which are used to discriminate signal events from the copious backgrounds~\cite{stange}.
In this Letter, we combine the searches from the CDF
and D0 Collaborations for $H$ bosons produced in association with a vector boson,
with subsequent decays $H\rightarrow b{\bar{b}}$.
Both collaborations consider the processes
\WH, \ZH, and \metbb~\cite{cdfwh2012,cdfzh2012,cdfzhll2012,dzwh2012,dzzh2012,dzzhll2012}
(where $\ell$ is either $e$ or $\mu$ and \met\ denotes missing transverse energy~\cite{coord}),
and separately combine results within their collaborations~\cite{cdfHbb2012,dzHbb2012}.
This is the first publication of a combination of CDF and D0's searches for $H\to b\bar{b}$,
which is based on the preliminary findings reported in Ref.~\cite{tevcomb2012}.

Much is known about the Higgs boson from other experiments.  The direct searches 
at LEP2 in the $e^+e^-\rightarrow ZH(\rightarrow b{\bar{b}})$ mode, with a small 
contribution from vector boson fusion, are very similar to those combined here, 
and exclude SM Higgs boson masses below 114.4~GeV/$c^2$ at the 95\% confidence 
level (C.L.)~\cite{sm-lep}.  Direct searches for $VH\rightarrow Vb{\bar{b}}$ at 
the LHC, where $V=W$ or $Z$~\cite{atlasbb,cmsbb}, do not yet constrain the allowed 
SM Higgs boson mass range.  Including other search modes, direct searches at the 
LHC for the SM Higgs boson limit its mass to be between 116.6 and 119.4~GeV/$c^2$
or between 122.1 and 127.0~GeV/$c^2$, at the 95\% C.L.~\cite{cmscomb2012,atlascomb2012}.
Within these searches, both LHC experiments observe local excesses above the background 
expectations for a Higgs boson mass of approximately 125~GeV/$c^2$.
With additional data and
analysis improvements, the LHC experiments confirm these excesses and observe
a particle with properties consistent with those of the SM Higgs boson~\cite{lhcobs}.
Much of the power of the LHC searches comes
from $gg\rightarrow H$ production and Higgs boson decays to $\gamma\gamma$, $W^+W^-$, and $ZZ$, which probe the
couplings of the Higgs boson to other bosons.  In the allowed mass range, the Tevatron experiments are particularly sensitive
to {\it VH} production with $H\rightarrow b{\bar{b}}$, which probes the Higgs boson's coupling to $b$ quarks.
We search for Higgs bosons of masses $100<m_H<150$~GeV/$c^2$ and interpret our results independently
of searches which are not sensitive to the specific Higgs boson production and decay modes studied here.
We also report results assuming $m_H=125$~GeV/$c^2$.

%These results are consistent with 
%precision electroweak data, including the recently updated measurements of the top quark mass $m_t$
%and the $W$-boson mass $M_W$ from
%the CDF and D0 Collaborations~\cite{topmass,wmass}, which yield an
%indirect constraint on the allowed mass of the Higgs boson, $m_H < 152$~GeV/$c^2$~\cite{elweak}
%at the 95\% C.L.  

Higgs boson signal events are simulated using the leading order (LO) calculation from 
\PYTHIA~\cite{pythia},
with CTEQ5L (CDF) and CTEQ6L1 (D0)~\cite{cteq} parton distribution functions (PDFs).
We normalize our Higgs boson signal-rate predictions to the highest-order calculations
available.  The {\it WH} and {\it ZH} cross section calculations are
performed at next-to-next-to leading order (NNLO) precision in QCD and
next-to-leading-order (NLO) precision in the electroweak corrections~\cite{vhtheory}.  We use the branching fractions
for Higgs boson decay from Ref.~\cite{lhcdifferential}.
These rely on calculations using \HDECAY~\cite{hdecay}
and {\sc prophecy4f}~\cite{prophecy4f}.
Assuming the $m_H$ = 125~GeV/$c^2$
hypothesis, we expect approximately 155 Higgs boson signal events to pass our
selection requirements, along with $9.2\times 10^4$ background events from all 
other SM sources.

We model SM and instrumental background processes using a mixture of Monte 
Carlo (MC) and data-driven methods.  For CDF, backgrounds from SM processes 
with electroweak gauge bosons or top quarks are modeled using \textsc{PYTHIA}, 
\textsc{ALPGEN}~\cite{Mangano:2002ea}, \textsc{MC@NLO}~\cite{MC@NLO}, and 
\textsc{HERWIG}~\cite{herwig}.  For D0, these backgrounds are modeled using 
\textsc{PYTHIA}, \textsc{ALPGEN}, and \textsc{COMPHEP}~\cite{comphep}.  An
interface to \textsc{PYTHIA} provides parton showering and hadronization for 
generators without this functionality.

Diboson ({\it{WW}}, {\it{WZ}}, {\it{ZZ}}) MC samples are normalized using 
the NLO calculations from \MCFM~\cite{mcfm}.  For $t{\bar{t}}$, we use a
production cross section of $7.04\pm 0.70$~pb~\cite{mochuwer}, which is
based on a top-quark mass of 173~GeV/$c^2$~\cite{topmass} and MSTW 2008 
NNLO PDFs~\cite{MSTW}.  The single-top-quark production cross section is 
taken to be $3.15\pm 0.31$~pb~\cite{kidonakis_st}.  Data-driven methods 
are used to normalize the {\it W/Z} plus light-flavor and heavy-flavor 
jet backgrounds~\cite{hfjet} using {\it W/Z} data events containing no
$b$-tagged jets~\cite{btagdef}, which have negligible signal content~\cite{cdfHbb2012,dzHbb2012}.

The CDF and D0 detectors are multipurpose solenoidal spectrometers surrounded by 
hermetic calorimeters and muon detectors and are designed to study the products
of 1.96 TeV proton-antiproton collisions~\cite{cdfdetector,d0detector}.
All searches combined here use the complete Tevatron data sample,
which after data quality requirements corresponds to 9.45 fb$^{-1}$~--~9.7~fb$^{-1}$; 
the size of the analyzed data set depends 
on the experiment and the search channel.
The online event selections (triggers) rely on fast reconstruction
of combinations of high-$p_T$ lepton candidates, jets, and \met{}.
Event selections are similar in the CDF and D0
analyses, consisting typically of a preselection based on event topology and kinematics,
and a subsequent selection using
$b$-tagging.  Both collaborations use multivariate analysis (MVA) techniques 
that combine several 
discriminating variables into a single final discriminant which is used to separate signal from background.  Each
channel is divided into exclusive sub-channels according to various
lepton, jet multiplicity, and $b$-tagging characterization criteria aimed at
grouping events with similar signal-to-background ratio and so optimize
the overall sensitivity.
%For CDF (D0) there are 16 (3), 26 (8), and 3 (2) sub-channels for the \ZH{}, \WH{},
%and \metbb{} analyses, respectively, totaling to 45 (13) for the
%combination presented here.  
Due to the importance of $b$-tagging,
both collaborations have developed multivariate approaches to maximize the performance
of the $b$-tagging algorithms.  
A boosted decision tree algorithm is used in the D0 analysis, which builds and improves
upon the previous neural network $b$-tagger~\cite{Abazov:2010ab},
giving an identification efficiency of
$\approx 80\%$ for $b$-jets with a mis-identification rate of
$\approx 10\%$.  The CDF $b$-tagging algorithm has been recently augmented
with an MVA~\cite{tagging}, providing a $b$-tagging efficiency
of $\approx 70$\% and a mis-identification rate of $\approx 5$\%.

The reconstructed
dijet mass provides discrimination between signal and
background.  The decay width of the Higgs boson is expected to be
much narrower than the experimental dijet mass resolution, which is
typically 15\% of the mean
reconstructed mass.
A SM Higgs boson signal would appear as a broad
enhancement in the reconstructed dijet mass distribution.  The sensitivity is
enhanced by combining the dijet mass with other kinematic information using
multivariate discriminants.  The MVA
functions are optimized separately for each sub-channel and for each
hypothesized value of $m_H$ in the range 100--150~GeV/$c^2$, in
5~GeV/$c^2$ intervals.  The results from each sub-channel are summarized
in histograms of the MVA discriminants for the expected Higgs boson signals,
the backgrounds itemized by source, and the observed data.

We interpret the results using
both Bayesian and modified frequentist techniques, separately at each value of $m_H$.
These methods are described in Refs.~\cite{tevcomb2012,tevwwprl,pdgstats}.  These techniques are
built on a likelihood function which is a product of Poisson probabilities for observing
the data in each bin of each sub-channel.  Systematic uncertainties are parametrized with
nuisance parameters, which affect the rates of the predicted 
signal and background yields in each bin.  A nuisance parameter may affect
the predictions of multiple sources of signal and background in multiple sub-channels,
thus taking correlations into account.  A nuisance parameter may also affect multiple bins' predictions by
different amounts, thus parameterizing uncertainty in the shapes of distributions.
Gaussian priors are assumed for the nuisance parameters,
truncated to ensure that no prediction is negative.  The signal predictions used 
correspond to SM Higgs boson production and decay, scaled by a factor $R$ for all 
bins of all sub-channels.  By scaling all signal contributions by the same factor, we assume that
the relative contributions of the different processes are as predicted by the SM.

In the Bayesian technique, we assume a uniform prior in $R$ and integrate the likelihood function
multiplied by the priors of the nuisance parameters to obtain the posterior density for $R$.
The observed 95\% credibility level upper limit on $R$, $R_{95}^{\rm{obs}}$, is such that 95\% of 
the integral of the posterior of $R$ is below $R_{95}^{\rm{obs}}$.  The expected distribution of 
$R_{95}$ is computed in an ensemble of simulated experimental outcomes assuming no signal is present.  
In each simulated outcome, random values of the nuisance parameters are drawn from their priors.  
A combined measurement of the cross section for Higgs boson production times the branching fraction 
${\cal B}(H\rightarrow b{\bar{b}})$, in units of the SM production rate, is given by $R^{\rm{fit}}$,
which is the value of $R$ that maximizes the posterior density. The 68\% credibility interval, which 
corresponds to one standard deviation (s.d.), is quoted as the smallest interval containing 68\% of 
the integral of the posterior.

We also perform calculations using the modified frequentist technique~\cite{pdgstats},
${\rm CL}_{\rm s}$, using
a log-likelihood ratio ($LLR$) as the test statistic:
\begin{equation}
LLR = -2\ln\frac{p({\mathrm{data}}|H_1)}{p({\mathrm{data}}|H_0)},
\end{equation}
where $H_1$ denotes the test hypothesis, which admits the presence of
SM backgrounds and a Higgs boson signal, $H_0$ denotes the null
hypothesis, for only SM backgrounds, and `data' are either simulated 
data constructed from the expected signal and backgrounds, or the
actual observed data.  The probabilities $p$ are
computed using the best-fit posterior values of the nuisance parameters for each
simulated experimental outcome, separately for each of the two hypotheses, and include the
Poisson probabilities of observing the data multiplied by Gaussian
constraint terms for the values of the nuisance parameters.
The  ${\rm CL}_{\rm s}$ technique involves computing two $p$-values,
\begin{equation}
{\rm CL}_{\rm b} = p(LLR\ge LLR_{\mathrm{obs}} | H_0),
\end{equation}
where $LLR_{\mathrm{obs}}$ is the value of the test statistic computed for the
data, and
\begin{equation}
{\rm CL}_{\rm s+b} = p(LLR\ge LLR_{\mathrm{obs}} | H_1).
\end{equation}
To compute limits, we use the ratio of $p$-values, ${\rm CL}_{\rm s}={\rm CL}_{\rm s+b}/{\rm CL}_{\rm b}$.
If ${\rm CL}_{\rm s}<0.05$ for a particular choice
of $H_1$, parametrized by the signal scale factor $R$,
that hypothesis is excluded at the 95\% C.L.  The median expected limit is computed
using the median $LLR$ value expected in the background-only hypothesis.

The uncertainties on the
signal production cross sections are estimated from the factorization
and renormalization scale variations, which include the impact of
uncalculated higher-order corrections, as well as uncertainties due to PDFs, and
the dependence on the strong coupling constant, $\alpha_s$. The
resulting uncertainties on the inclusive {\it WH} and {\it ZH}
production rates are 7\%~\cite{vhtheory}.  We assign uncertainties to
the prediction of ${\cal B}(H\rightarrow b{\bar{b}})$ as calculated in
Ref.~\cite{denner11}.  These uncertainties arise from imperfect
knowledge of the mass of the $b$ and $c$ quarks, $\alpha_s$, and
theoretical uncertainties in the $b{\bar{b}}$ and $W^+W^-$ decay rates. 

The largest sources of uncertainty on the dominant backgrounds
are the rates of tagged $V$+heavy flavor jets, 
which are typically 20-30\% of the predicted values.  The posterior
uncertainties on these rates are typically 8\% or less.
Uncertainties on lepton identification and trigger efficiencies range
from 2\% to 6\% and are applied to both signal- and MC-based
background predictions.  These uncertainties are estimated from data-based
methods separately by CDF and D0, and differ based on lepton flavor and
identification category.  The $b$-tag efficiencies and mistag rates are similarly
constrained by auxiliary data samples, such as inclusive jet data or
$t{\bar{t}}$ events.  The uncertainty on the per-jet $b$-tag efficiency is
approximately 4\%, and the mistag uncertainties vary between 7\% and 15\%. 
The uncertainties on the measurements of the integrated luminosities, which
are used to normalize the expected signal yields and the MC-based 
backgrounds, are 6\% (CDF)~\cite{incdflumi}
and 6.1\% (D0)~\cite{ind0lumi}.  Of these values, 4\% arises from 
the inelastic $p{\bar{p}}$ cross section, which is
taken to be correlated between CDF and D0.

 \begin{figure}[htb] \begin{centering}
 \includegraphics[width=0.8\columnwidth]{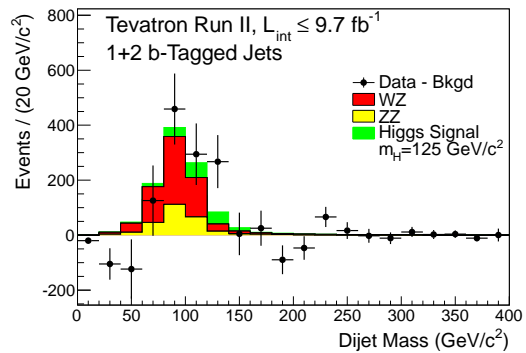}
\caption{
 \label{fig:bgsubmjj} Background-subtracted distribution of the reconstructed dijet mass $m_{jj}$, 
 summed over all input channels.  The {\it VZ} signal and the background contributions are fit 
 to the data, and the fitted background is subtracted.  The fitted {\it VZ} and expected SM Higgs 
 ($m_H=125$~GeV/$c^2$) contributions are shown with filled histograms.}
 \end{centering}
 \end{figure}

 \begin{figure}[htb] \begin{centering}
 \includegraphics[width=0.8\columnwidth]{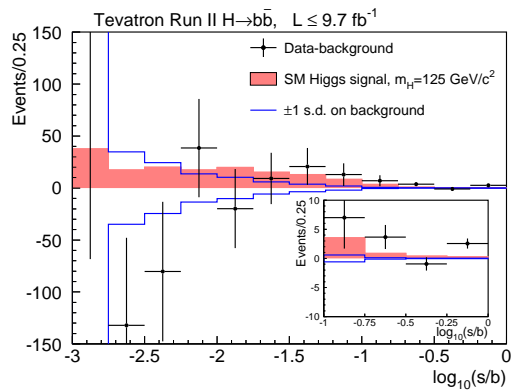}
\caption{
 \label{fig:bgsub} Background-subtracted
 distribution for the 
 discriminant histograms, summed for bins with similar
 signal-to-background ratio ($s/b$), for
 the $H\rightarrow b{\bar{b}}$ $(m_H=125$~GeV/$c^2)$ search.
The solid histogram shows the uncertainty on the background after the fit
to the data as discussed in the text.
 The signal model, scaled to the SM expectation, is shown with a 
 filled histogram.  Uncertainties on the data points correspond 
 to the square root of the sum of the expected signal and background 
 yields in each bin.}
 \end{centering}
 \end{figure}

\begin{figure}[hb]
\begin{centering}
\includegraphics[width=0.8\columnwidth]{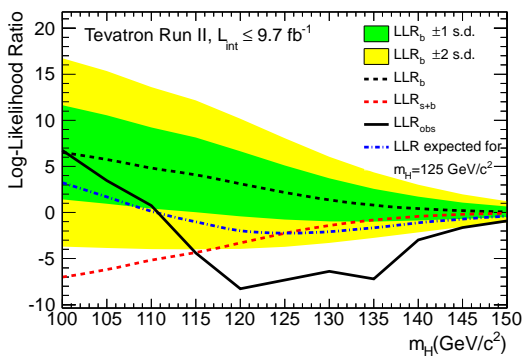}
\caption{
\label{fig:llr}
The log-likelihood ratio $LLR$ as a function of Higgs boson mass.
The dark and light-shaded bands correspond to the regions encompassing 1~s.d.~and 2~s.d.~fluctuations 
of the background, respectively.   The dot-dashed line shows the median expected $LLR$ assuming the 
SM Higgs boson is present at $m_H=125$~GeV/$c^2$.}
\end{centering}
\end{figure}

\begin{figure}[t]
\begin{centering}
\includegraphics[width=7.4cm]{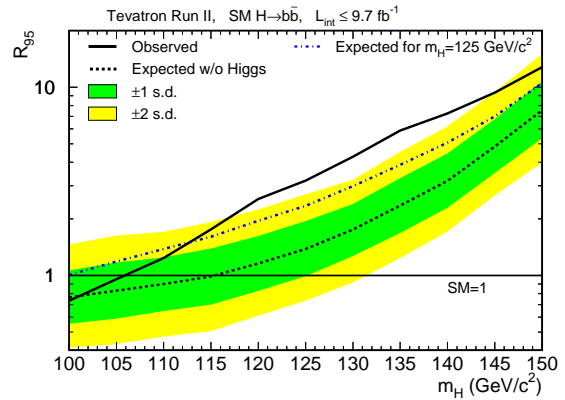}
\caption{
\label{fig:limits}
The observed 95\% credibility level upper limits on SM Higgs boson
production ($R_{95}$) as a function of Higgs boson mass.  The dashed 
line indicates the median expected value in the absence of a signal, 
and the shaded bands indicate the 1 s.d.~and 2 s.d.~ranges in which 
$R_{95}$ is expected to fluctuate.  The dot-dashed line shows the 
median expected limit if the SM Higgs boson is present at $m_H=125$~GeV/$c^2$.}
\end{centering}
\end{figure}

 \begin{figure}[htb] \begin{centering}
 \includegraphics[width=7.4cm]{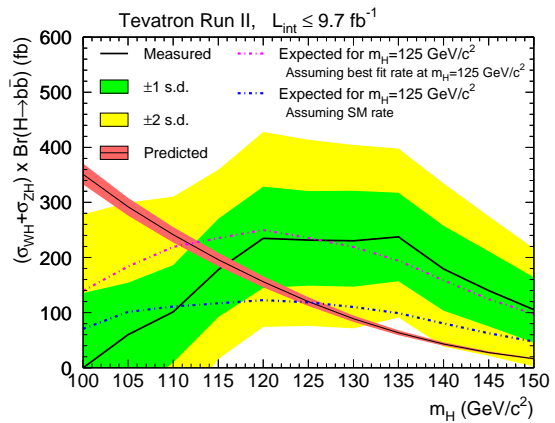} \caption{
 \label{fig:xsfit} The best-fit cross section times branching ratio
 $(\sigma_{WH}+\sigma_{ZH})\times \mathcal{B}(H\rightarrow
 b{\bar{b}})$ as a function of $m_H$.  The dark and light-shaded 
 regions indicate the 1 s.d.~and 2 s.d.~measurement uncertainties, 
 and the SM prediction is
 shown as the smooth, falling curve with a narrow band indicating 
 the theoretical uncertainty.   The expected cross section fit values 
 assuming the SM Higgs boson is present at $m_H=125$~GeV/$c^2$ are 
 shown with dot-dashed lines for the cases of the expected SM rate 
 (dark blue) and the best fitted rate from data (light magenta).} \end{centering} \end{figure}

 \begin{figure}[htb] \begin{centering}
 \includegraphics[width=7.4cm]{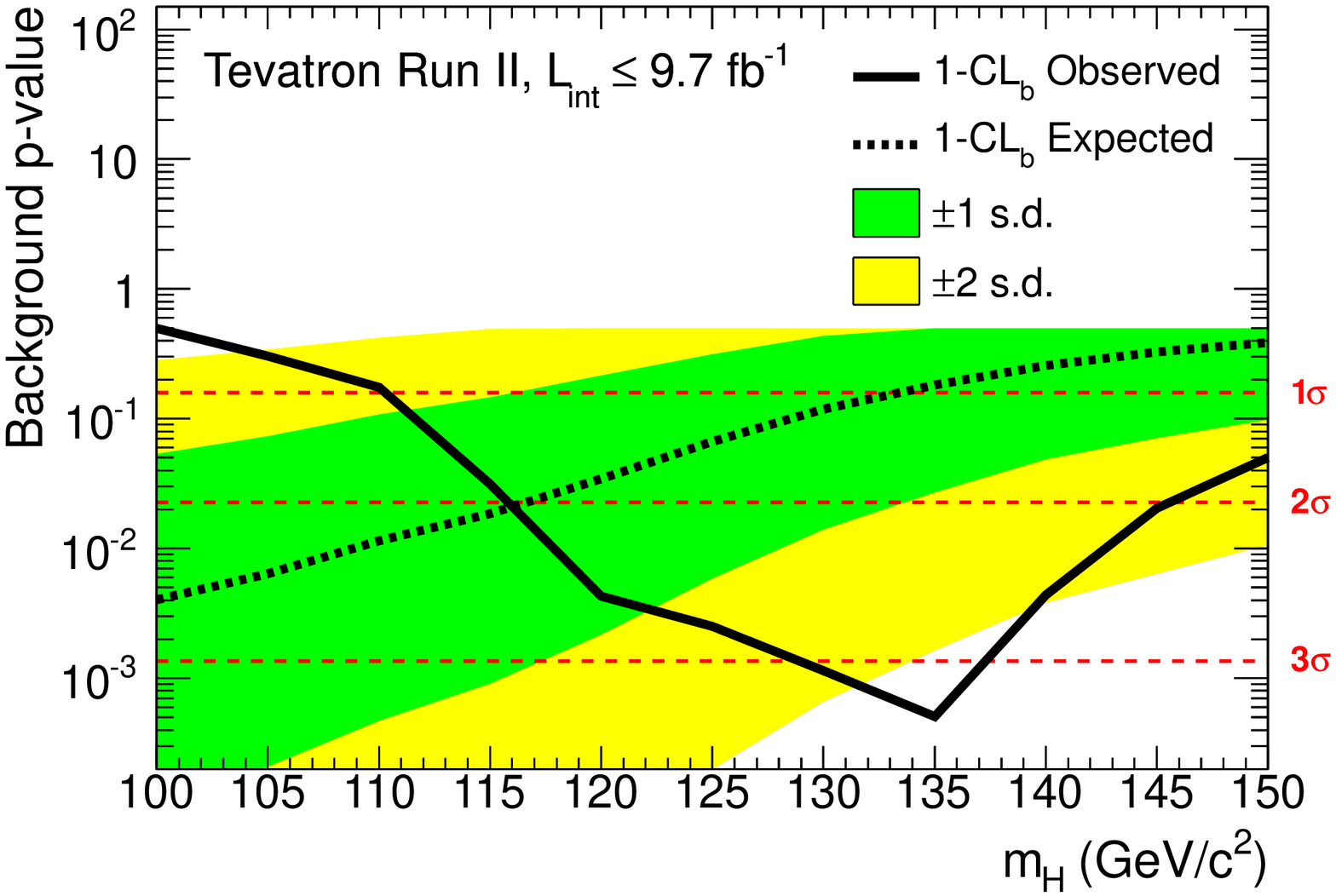} \caption{
 \label{fig:pvalue} The $p$-value as a function of $m_H$ under the 
 background-only hypothesis.  Also shown are the median expected 
 values assuming a SM signal is present, evaluated separately at 
 each $m_H$.  The associated dark and light-shaded bands indicate 
 the 1 s.d.~and 2 s.d.~fluctuations of possible experimental 
 outcomes.}  
 \end{centering} \end{figure}

To validate our background modeling and search methods, we
perform a search for SM diboson production in the same final 
states used for the SM $H\rar b\bar{b}$ searches.  The NLO SM
cross section for {\it VZ} times the branching fraction of $Z\rar 
b\bar{b}$ is 0.68 $\pm$ 0.05~pb, which is about six times larger 
than the 0.12 $\pm$ 0.01~pb cross section times branching fraction 
of $VH(H\rar b\bar{b})$ for a 125~GeV/$c^2$ SM Higgs boson.  The 
data sample, reconstruction, process modeling, uncertainties, and 
sub-channel divisions are identical to those of the SM Higgs boson 
search.  However, discriminant functions are trained to distinguish 
the contributions of SM diboson production from those of other 
backgrounds, and potential contributions from Higgs boson production
are not considered.  The measured cross section for {\it VZ} is 
$3.9 \pm 0.6$~(stat)~$\pm 0.7$~(syst)~pb, which is consistent with 
the SM prediction of $4.4 \pm 0.3$~pb. 

The combined background-subtracted reconstructed dijet mass ($m_{jj}$) 
distribution for the {\it VZ} analysis is shown in Fig.~\ref{fig:bgsubmjj}.  
The {\it VZ} signal and the background contributions are fit to the data, 
and the fitted background is subtracted.  Also shown is the contribution 
expected from a SM Higgs boson with $m_H=125$~GeV/$c^2$.  

To visualize the results produced by the 
multivariate {\it VH} analyses, we combine the histograms of the final
discriminants, adding the contents of bins with similar
signal-to-background ratio ($s/b$).  Figure~\ref{fig:bgsub} shows the
signal expectation and the data with the background (including {\it VZ}) subtracted, as a
function of the $s/b$ of the collected bins, for the combined Higgs boson search, assuming
$m_H=125$~GeV/$c^2$.  The background model is fit to the data,
and the uncertainties on the background are those after the nuisance
parameters have been constrained in the fit.  An excess of 
events in the highest $s/b$ bins relative to the
background-only expectation is observed.
We also show the $LLR$ as a function of $m_H$ in Fig.~\ref{fig:llr}, along with its
expected values under the hypotheses $H_0$ and $H_1$, and also the hypothesis that a SM Higgs boson
is present with $m_H=125$~GeV/$c^2$.

We extract limits on SM Higgs boson production as a function of $m_H$ in the range
100--150~GeV/$c^2$ in terms of $R_{95}^{\mathrm{obs}}$, the observed limit relative 
to the SM rate.
These limits are shown in Fig.~\ref{fig:limits},
together with the median expected values and distributions in
simulated experimental outcomes assuming a signal is absent.  We also show the median expected limits assuming the SM Higgs boson,
with $m_H=125$~GeV/$c^2$, is present.
We exclude $m_H< 106$~GeV/$c^2$ at the 95\% credibility level,
while our median expected limit on $m_H$ is 116~GeV/$c^2$, if no signal were present.
The exclusions obtained with the ${\rm CL}_{\rm s}$ technique match those computed with the
Bayesian technique.

The observed limits are weaker than expected due to an excess events in the data with
respect to the background predictions in the most sensitive bins of the discriminant distributions, favoring
the hypothesis that a signal is present.  We characterize this excess by
computing the best-fit rate parameter
$R^{\rm{fit}}$, which, when multiplied by the SM prediction for the
associated production cross section times the decay branching ratio
($\sigma_{WH}+\sigma_{ZH})\times \mathcal{B}(H\rightarrow
b{\bar{b}})$, yields the best fit value for this quantity. We show our
fitted ($\sigma_{WH}+\sigma_{ZH})\times \mathcal{B}(H\rightarrow
b{\bar{b}})$ as a function of $m_H$, along with the SM prediction, in
Fig.~\ref{fig:xsfit}.  The figure also shows the expected cross section 
fits for each $m_H$ assuming that the SM Higgs boson, with $m_H=125$~GeV/$c^2$, 
is present.  The expected fits are shown for both the expected SM rate 
and the best fitted rate from data, which corresponds to 
%an associated production cross section
%times the decay branching ratio of 
$(\sigma_{WH}+\sigma_{ZH})\times
\mathcal{B}(H\rightarrow b{\bar{b}})= 0.23 ^{+0.09}_{-0.08}~(\mathrm{stat+syst})$~pb.
%For a mass hypothesis of $m_H=125$~GeV/$c^2$, we measure 
The corresponding SM prediction for $m_H=125$~GeV/$c^2$ is 0.12 $\pm$ 0.01~pb.

The significance of the excess in the data over the background prediction is
computed at each hypothesized Higgs boson mass in the range 100--150~GeV/$c^2$
by calculating the local $p$-value under the background-only hypothesis using
$R^{\rm{fit}}$ as the test statistic.  This $p$-value expresses the probability 
to obtain the value of $R^{\rm{fit}}$ observed in the data or larger, assuming 
a signal is truly absent.  These $p$-values are shown in Fig.~\ref{fig:pvalue} 
along with the expected $p$-values assuming a SM signal is present, separately 
for each value of $m_H$.  The observed $p$-value as a function of $m_H$ exhibits
a broad minimum and the maximum local significance corresponds to 3.3 
standard deviations at $m_H=135$~GeV/$c^2$.

The Look-Elsewhere Effect (LEE)~\cite{LEE,Dunn1961Multiple} accounts for the 
possibility of a background fluctuation affecting the local $p$-value anywhere 
in the tested $m_H$ range.  In the mass range from 115~GeV/$c^2$ (the prior 
bound from the LEP2 direct search~\cite{sm-lep}) to 150~GeV/$c^2$, the
reconstructed mass resolution is typically $15$\%, and the resulting LEE factor 
is approximately $2$.
Correcting for the LEE yields a global significance of 3.1 standard deviations.  
Taking into account the exclusion limits for the SM Higgs boson mentioned 
earlier, there is no LEE and we derive a significance of 2.8 standard deviations 
for $m_H=125$~GeV/$c^2$.

We interpret this result as evidence for the presence of a particle that is 
produced in association with a $W$ or $Z$ boson and decays to a bottom-antibottom 
quark pair.  The excess seen in the data is most significant in the mass range 
between 120 and 135~GeV/$c^2$, and is consistent with production of the SM Higgs 
boson within this mass range.  Assuming a Higgs boson exists in this mass range,
these results provide a direct probe of its coupling to $b$ quarks.

\begin{center}
{\bf Acknowledgements}
\end{center}

% acknowledgement.tex                            13 May 2012 
%
We thank the Fermilab staff and technical staffs of the participating institutions for their vital contributions and acknowledge support from the
DOE and NSF (USA),
ARC (Australia),
CNPq, FAPERJ, FAPESP and FUNDUNESP (Brazil),
NSERC (Canada),
NSC, CAS and CNSF (China),
Colciencias (Colombia),
MSMT and GACR (Czech Republic),
the Academy of Finland,
CEA and CNRS/IN2P3 (France),
BMBF and DFG (Germany),
DAE and DST (India),
SFI (Ireland),
INFN (Italy),
MEXT (Japan),
the Korean World Class University Program and NRF (Korea),
CONACyT (Mexico),
FOM (Netherlands),
MON, NRC KI and RFBR (Russia),
the Slovak R\&D Agency, 
the Ministerio de Ciencia e Innovaci\'{o}n, and Programa Consolider-Ingenio 2010 (Spain),
The Swedish Research Council (Sweden),
SNSF (Switzerland),
STFC and the Royal Society (United Kingdom),
and the A.P Sloan Foundation (USA).

%------------------------------------------------------------------------------%       Bibliography
\bibliographystyle{apsrev}

\begin{thebibliography}{46}
\expandafter\ifx\csname natexlab\endcsname\relax\def\natexlab#1{#1}\fi
\expandafter\ifx\csname bibnamefont\endcsname\relax
  \def\bibnamefont#1{#1}\fi
\expandafter\ifx\csname bibfnamefont\endcsname\relax
  \def\bibfnamefont#1{#1}\fi
\expandafter\ifx\csname citenamefont\endcsname\relax
  \def\citenamefont#1{#1}\fi
\expandafter\ifx\csname url\endcsname\relax
  \def\url#1{\texttt{#1}}\fi
\expandafter\ifx\csname urlprefix\endcsname\relax\def\urlprefix{URL }\fi
\providecommand{\bibinfo}[2]{#2}
\providecommand{\eprint}[2][]{\url{#2}}

\bibitem[{gws()}]{gws}
\bibinfo{note}{S.~L.~Glashow, Nucl.\ Phys.\ {\bf 22}, 579 (1961); \\
  S.~Weinberg, Phys. Rev. Lett. {\bf 19}, 1264 (1967); \\ A.~Salam, {\it
  Elementary Particle Theory}, ed. N.~Svartholm (Almqvist and Wiksell,
  Stockholm), 367 (1968).}

\bibitem[{hig()}]{higgs}
\bibinfo{note}{F.~Englert and R.~Brout, Phys. Rev. Lett. {\bf 13}, 321 (1964);
  \\ P.~W. Higgs, Phys. Rev. Lett. {\bf 13}, 508 (1964); \\ G.~S.~Guralnik,
  C.~R.~Hagen, and T.~W.~B.~Kibble, Phys. Rev. Lett. {\bf 13}, 585 (1964); \\
  P.~W. Higgs, Phys. Rev. {\bf 145}, 1156 (1966).}

\bibitem[{vht()}]{vhtheory}
\bibinfo{note}{J.~Baglio and A.~Djouadi, J. High Energy Phys. 10 (2010) 064;
  O.~Brein, R.~V.~Harlander, M.~Weisemann, and T.~Zirke, Eur. Phys. J. C {\bf
  72}, 1868 (2012).}

\bibitem[{lhc({\natexlab{a}})}]{lhcdifferential}
\bibinfo{note}{S.~Dittmaier {\it et al.} (LHC Higgs Cross Section Working
  Group), arXiv:1201.3084 (2012).}

\bibitem[{sta()}]{stange}
\bibinfo{note}{A.~Stange, W.~Marciano, and S.~Willenbrock, Phys. Rev. D {\bf
  49}, 1354 (1994); A.~Stange, W.~Marciano, and S.~Willenbrock, Phys. Rev. D
  {\bf 50}, 4491 (1994).}

\bibitem[{cdf({\natexlab{a}})}]{cdfwh2012}
\bibinfo{note}{T. Aaltonen {\it et al.} (CDF Collaboration), arXiv:1207.1703,
  submitted to Phys. Rev. Lett.}

\bibitem[{cdf({\natexlab{b}})}]{cdfzh2012}
\bibinfo{note}{T. Aaltonen {\it et al.} (CDF Collaboration), arXiv:1207.1711,
  submitted to Phys. Rev. Lett.}

\bibitem[{cdf({\natexlab{c}})}]{cdfzhll2012}
\bibinfo{note}{T. Aaltonen {\it et al.} (CDF Collaboration), arXiv:1207.1704,
  submitted to Phys. Rev. Lett.}

\bibitem[{dzw()}]{dzwh2012}
\bibinfo{note}{V.M. Abazov {\it et al.} (D0 Collaboration),
  arXiv:1208.0653, submitted to Phys. Rev. Lett.}
%  FERMILAB-PUB-12-405-E, to be submitted to Phys. Rev. Lett.}

\bibitem[{dzz({\natexlab{a}})}]{dzzh2012}
\bibinfo{note}{V.M. Abazov {\it et al.} (D0 Collaboration), arXiv:1207.5689
  [hep-ex], submitted to Phys. Lett. B.}

\bibitem[{dzz({\natexlab{b}})}]{dzzhll2012}
\bibinfo{note}{V.M. Abazov {\it et al.} (D0 Collaboration), arXiv:1207.5819
  [hep-ex], submitted to Phys. Rev. Lett.}

\bibitem[{coo()}]{coord}
\bibinfo{note}{CDF and D0 use cylindrical coordinate systems with origins in
  the centers of the detectors, where $\theta$ and $\phi$ are the polar and
  azimuthal angles, respectively, and pseudorapidity is $\eta= - \rm ln$ $\rm
  tan(\theta/2)$.  The transverse energy, as measured by the calorimetry, is
  defined to be $E_T=E\sin\theta$.
  The missing $E_T$ (\METVEC) is defined by \METVEC$ =
  -\sum_iE_T^i{\hat n}_i, i={\rm calorimeter~tower~number}$, where ${\hat n}_i$
  is a unit vector perpendicular to the beam axis and pointing at the $i$th
  calorimeter tower. ~\METVEC{} is corrected for high-energy muons and also jet
  energy corrections. We define \MET=$|$\METVEC$|$. The transverse momentum
  $p_T$ is defined to be $p\sin\theta$.}

\bibitem[{cdf({\natexlab{d}})}]{cdfHbb2012}
\bibinfo{note}{T. Aaltonen {\it et al.} (CDF Collaboration), arXiv:1207.1707,
  submitted to Phys. Rev. Lett.}

\bibitem[{dzH()}]{dzHbb2012}
\bibinfo{note}{V.M. Abazov {\it et al.} (D0 Collaboration),
  arXiv:1207.6631, submitted to Phys. Rev. Lett.}

\bibitem[{tev({\natexlab{a}})}]{tevcomb2012}
\bibinfo{note}{The CDF and D0 Collaborations and the Tevatron New Physics and
  Higgs Working Group, arXiv:1207.0449 (2012).}

\bibitem[{sm-()}]{sm-lep}
\bibinfo{note}{The ALEPH, DELPHI, L3 and OPAL Collaborations, and the LEP
  Working Group for Higgs Boson Searches, Phys.\ Lett. B {\bf 565}, 61 (2003).}

\bibitem[{atl({\natexlab{a}})}]{atlasbb}
\bibinfo{note}{G. Aad {\it et al.} (ATLAS Collaboration), arXiv:1207.0210 (2012), submitted to Phys.
  Lett. B.}

\bibitem[{cms({\natexlab{a}})}]{cmsbb}
\bibinfo{note}{S.~Chatrchyan {\it et al.} (CMS Collaboration), Phys.\ Lett.\ B
  {\bf 710}, 284 (2012).}

\bibitem[{cms({\natexlab{b}})}]{cmscomb2012}
\bibinfo{note}{S.~Chatrchyan {\it et al.} (CMS Collaboration), Phys.\ Lett.\ B
  {\bf 710}, 26 (2012).}

\bibitem[{atl({\natexlab{b}})}]{atlascomb2012}
\bibinfo{note}{G.~Aad {\it et al.}, (ATLAS Collaboration), arXiv:1207.0319
  (2012), accepted by Phys. Rev. D.}

\bibitem[{lhc({\natexlab{b}})}]{lhcobs}
\bibinfo{note}{G.~Aad {\it et al.} (ATLAS Collaboration), arXiv:1207.7214 (2012), submitted to Phys. Lett. B;
S.~Chatrchyan {\it et al.} (CMS Collaboration), arXiv:1207.7235 (2012), submitted to Phys. Lett. B.}

\bibitem[{pyt()}]{pythia}
\bibinfo{note}{T.~Sj\"ostrand, S.~Mrenna, and P.~Skands, J.~High Energy Phys. 05
  (2006) 026. We use \PYTHIA{}~version 6.216 to generate the Higgs boson
  signals.}

\bibitem[{cte()}]{cteq}
\bibinfo{note}{H.~L.~Lai {\it et al.}, Eur. Phys. J. C~{\bf 12}, 375 (2000);
  J.~Pumplin {\it et al.}, JHEP {\bf 0207}, 012 (2002).}

\bibitem[{hde()}]{hdecay}
\bibinfo{note}{A.~Djouadi, J.~Kalinowski, and M.~Spira, Comput.\ Phys.\
  Commun.\ {\bf 108}, 56 (1998).}

\bibitem[{pro()}]{prophecy4f}
\bibinfo{note}{A.~Bredenstein, A.~Denner, S.~Dittmaier, and M.~M.~Weber, Phys.
  Rev. D {\bf 74}, 013004 (2006); A. Bredenstein, A.~Denner, S.~Dittmaier, A.
  M\"uck, and M.~M.~Weber, J. High Energy Phys. 02 (2007) 080.}

\bibitem[{Man()}]{Mangano:2002ea}
\bibinfo{note}{M.~Mangano, M.~Moretti, F.~Piccinini, R.~Pittau, and A.~Polosa,
  J. High Energy Phys. 07 (2003) 001.}

\bibitem[{MC@()}]{MC@NLO}
\bibinfo{note}{S. Frixione and B.R. Webber, J. High Energy Phys. {\bf 0206},
  029 (2002).}

\bibitem[{her()}]{herwig}
\bibinfo{note}{G.~Corcella {\it et al.}, J. High Energy Phys. {\bf 0101}, 010
  (2001).}

\bibitem[{com()}]{comphep}
\bibinfo{note}{A.~Pukhov {\it et al.}, arXiv:hep-ph/9908288 (1999); E.~Boos
  {\it et al.}, Nucl. Instrum. Methods A {\bf 534}, 250 (2004); E. Boos {\it et
  al.}, Phys. Atom. Nucl. {\bf 69}, 1317 (2006).}

\bibitem[{mcf()}]{mcfm}
\bibinfo{note}{J.~M.~Campbell and R.~K.~Ellis, Phys.\ Rev.\ D {\bf 60}, 113006
  (1999).}

\bibitem[{moc()}]{mochuwer}
\bibinfo{note}{U.~Langenfeld, S.~Moch, and P.~Uwer, Phys. Rev. D {\bf 80},
  054009 (2009).}

\bibitem[{top()}]{topmass}
\bibinfo{note}{T. Aaltonen {\it et. al.} (CDF and D0 Collaborations),
  arXiv:1207.1069 (2012), submitted to Phys. Rev. D.}

\bibitem[{MST()}]{MSTW}
\bibinfo{note}{A.~D.~Martin, W.~J.~Stirling, R.~S.~Thorne and G.~Watt, Eur.\
  Phys.\ J.\ C {\bf 63}, 189 (2009).}

\bibitem[{kid()}]{kidonakis_st}
\bibinfo{note}{N.~Kidonakis, Phys.\ Rev.\ D {\bf 74}, 114012 (2006).}

\bibitem[{hfj()}]{hfjet}
\bibinfo{note}{A heavy-flavor jet is a reconstucted cluster of calorimeter
  energies associated with particles produced in the hadronization and decay of
  a bottom or charm quark.}

\bibitem[{bta()}]{btagdef}
\bibinfo{note}{A $b$-tagged jet is one identified as consistent with that
  expected from the decay products of a bottom quark based on properties such
  as the presence of displaced track vertices or soft leptons.}

\bibitem[{cdf({\natexlab{e}})}]{cdfdetector}
\bibinfo{note}{D.~Acosta, {\it et al.} (CDF Collaboration), Phys. Rev. D {\bf
  71}, 032001 (2005); A.~Abulencia, {\it et al.} (CDF Collaboration), J. Phys.
  G Nucl. Part. Phys. {\bf 34}, 2457 (2007).}

\bibitem[{d0d()}]{d0detector}
\bibinfo{note}{V.~M.~Abazov {\it et al.} (D0 Collaboration), Nucl. Instrum.
  Methods Phys. Res. A {\bf 565}, 463 (2006); M.~Abolins {\it et al.}, Nucl.
  Instrum. Methods A {\bf 584}, 75 (2008); R.~Angstadt {\it et al.}, Nucl.
  Instrum. Methods A {\bf 622}, 298 (2010).}

\bibitem[{Aba()}]{Abazov:2010ab}
\bibinfo{note}{V.~M.~Abazov {\it et al.}, Nucl.\ Instrum.\ Methods\ A {\bf
  620}, 490 (2010).}

\bibitem[{tag()}]{tagging}
\bibinfo{note}{J.~Freeman {\it et al.}, arXiv:1205.1812 (2012); D.~Acosta {\it
  et al.} (CDF Collaboration), Phys.\ Rev.\ D {\bf 71}, 052003 (2005);
  A.~Abulencia {\it et al.} (CDF Collaboration), Phys.\ Rev.\ D {\bf 74},
  072006 (2006).}

\bibitem[{tev({\natexlab{b}})}]{tevwwprl}
\bibinfo{note}{T. Aaltonen {\it et al.} (CDF and D0 Collaborations), Phys. Rev.
  Lett. {\bf 104}, 061802 (2010).}

\bibitem[{pdg()}]{pdgstats}
\bibinfo{note}{{\it Statistics}, in K. Nakamura et al. (Particle Data Group),
  J. Phys. G {\bf 37}, 075021 (2010); A.~L.~Read, J. Phys. G {\bf 28}, 2693
  (2002); W.~Fisher, ``Systematics and Limit Calculations,''
  FERMILAB-TM-2386-E.}

\bibitem[{den()}]{denner11}
\bibinfo{note}{A.~Denner, S.~Heinemeyer, I.~Puljak, D.~Rebuzzi, and M.~Spira,
  Eur. Phys. J. C {\bf 71}, 1753 (2011).}

\bibitem[{inc()}]{incdflumi}
\bibinfo{note}{S.~Klimenko, J.~Konigsberg, and T.~M.~Liss, FERMILAB-FN-0741
  (2003).}

\bibitem[{ind()}]{ind0lumi}
\bibinfo{note}{T. Andeen {\it et al.}, FERMILAB-TM-2365 (2007).}

\bibitem[{LEE()}]{LEE}
\bibinfo{note}{L.~Lyons, The Annals of Applied Statistics, Vol. 2, No. 3, 887
  (2008).}

\bibitem[{\citenamefont{Dunn}(1961)}]{Dunn1961Multiple}
\bibinfo{author}{\bibfnamefont{O.~J.} \bibnamefont{Dunn}},
  \bibinfo{journal}{Journal of the American Statistical Association}
  \textbf{\bibinfo{volume}{56}}, \bibinfo{pages}{52} (\bibinfo{year}{1961}).

\end{thebibliography}

\end{document}